\documentclass[prb,twocolumn,superscriptaddress,showpacs]{revtex4-2}
\usepackage{graphicx}% Include figure files
\usepackage{epsfig}% Include figure files
\usepackage{amssymb,amsmath,amsfonts,hyperref,bm}
\usepackage{wasysym}
\usepackage{latexsym}
\usepackage{eucal}
\usepackage{verbatim}
\usepackage[normalem]{ulem}
\usepackage{color}
\usepackage{cleveref}
\usepackage{physics}
\usepackage[final]{showkeys}
\usepackage[dvipsnames]{xcolor}
\hypersetup{urlcolor=RoyalBlue, colorlinks=true,citecolor=RoyalBlue,linkcolor=OrangeRed}

\newcommand{\avg}[1]{\left< #1 \right>} % for average

\newcommand{\be}{\begin{eqnarray}}
\newcommand{\ee}{\end{eqnarray}}

\begin{document}
\title{Fractional Brownian motion of worms in worm algorithms for frustrated Ising magnets}
\author{Geet Rakala}
\affiliation{\small{Okinawa Institute of Science and Technology Graduate University, Onna-son, Okinawa 904-0412, Japan}}
\author{Kedar Damle}
\affiliation{\small{Tata Institute of Fundamental Research, 1 Homi Bhabha Road, Mumbai 400005, India}}
\author{Deepak Dhar}
\affiliation{\small{Indian Institute of Science Education and Research, Homi Bhabha Road, Pashan, Pune 411008, India}} 

\begin{abstract}
	
  We study the distribution of lengths and other statistical properties of worms constructed by Monte Carlo worm algorithms in the power-law three-sublattice ordered phase of frustrated triangular and kagome lattice  Ising antiferromagnets. Viewing each step of the worm construction as a position increment (step) of a random walker, we demonstrate that the
  persistence exponent $\theta$ and the dynamical exponent $z$ of this random walk depend only on the universal power-law exponents of the underlying critical phase, and not on the details of the worm algorithm or the microscopic Hamiltonian.  Further, we argue that the detailed balance criterion obeyed by such worm algorithms and the power-law correlations of the underlying equilibrium system together give rise to two related properties of this random walk: First, the steps of the walk are expected to be power-law correlated in time.  Second, the position distribution of the walker relative to its starting point is given by the equilibrium position distribution of a particle in an attractive logarithmic central potential of strength $\eta_m$, where $\eta_m$ is the universal power-law exponent of the equilibrium defect-antidefect correlation function of the underlying spin system. We derive a scaling relation, $z = (2-\eta_m)/(1-\theta)$, that allows us to express
  the dynamical exponent $z(\eta_m)$ of this process in terms of its persistence exponent $\theta(\eta_m)$. Our measurements of $z(\eta_m)$ and $\theta(\eta_m)$ are consistent with this relation over
  a range of values of the universal equilibrium exponent $\eta_m$, and yield subdiffusive ($z>2$) values of $z$ in the entire range.
  Thus we demonstrate that
  the worms represent a discrete-time realization of a fractional Brownian motion characterized by these properties. 
\end{abstract}

\vskip2pc

\maketitle
\section{Introduction}

Worm algorithms are very useful as a means of generating non-local updates in Monte Carlo simulations of various lattice models (for a brief review, see Section 5.1 of Ref.~\onlinecite{Evertz}). The `worm' construction typically starts by creating a defect and an antidefect next to each other in the initial configuration. The defect and anti-defect take the system out of the configuration space of the physical system. The location of the defect defines the fixed
{\it tail} of the worm, while the {\it head} of the worm corresponds to the antidefect, which is propagated through the lattice in a way which satisfies detailed balance conditions in a larger configuration space that allows for one defect-antidefect pair. The construction ends when the {\it head} reaches the {\it tail} again and annihilates it. All physical variables encountered during the motion of the worm are updated as a result of this construction. 

An early implementation of a worm algorithm in the context of classical Monte Carlo simulations used the high-temperature expansion representation, and updated closed path configurations that represent terms in this expansion.\cite{Prokofev_Svistunov} A similar idea was also used to develop a worm algorithm for the quantum rotor model in $d=2$ spatial dimensions using the link-current representation (divergence-free configurations of current variables on links of the equivalent classical $d+1=3$ dimensional space-time lattice).\cite{Alet_Sorensen,Alet_Sorensen_1} The construction creates a charged defect (with nonzero divergence of the link current) at the tail, and a corresponding antidefect at the head. In this case, the worm maintains detailed balance in the configuration space relevant to the sampling of the single-particle Green's function of the system.\cite{Wallin_Sorensen_Girvin_Young} In quantum Monte Carlo simulations of other bosonic systems, a similar worm algorithm has been used both in the framework of imaginary time worldline formulations,\cite{Suzuki,Hirsch_et_al} and the stochastic series expansion (SSE) approach\cite{Sandvik} to perform non-local changes in the configuration. In this case too, the defects at the head and the tail of the worm correspond to creation and annihilation of a particle,\cite{Dorneich_Troyer} allowing access to configurations relevant to the sampling of the single-particle Green's function.

`Dual' worm algorithms have also been used to construct cluster updates for two-dimensional classical Ising models.\cite{Hitchcock_Sorensen_Alet} These algorithms work by updating dimer configurations (which encode bond energies of the original model) along a closed loop on the corresponding dual lattice. The updated bond energies are used to obtain a new spin configuration in which all spins in the interior of this closed loop have been flipped in one step. Recently, this approach has been used\cite{loopalgo} to obtain efficient cluster updates for frustrated Ising models for which the usual cluster updates\cite{Swendsen_Wang,Wolff} are known to perform poorly.\cite{Leung_Henley} For the antiferromagnetic Ising model on the triangular lattice, bond-energy configurations correspond to dimer configuratons on the dual honeycomb lattice, with dimers intersecting frustrated bonds on the direct lattice.  At $T=0$, ground states of the antiferromagnetic Ising model are characterized by the constraint that there is exactly one frustrated bond per triangle. The corresponding dual-lattice dimer configuration is hence {\em fully-packed} (every dual-lattice site is touched by a dimer) and characterized by a  {\em hard-core} constraint (dimers don't touch each other). At $T>0$ the dimer configurations do not obey a hard-core constraint, since configurations in which some dual-lattice sites have three dimers touching them (corresponding to triangles with three frustrated bonds) are accessed by thermal fluctuations; each dual-lattice site is now touched by one or three dimers. In the frustrated Ising antiferromagnet on the kagome lattice, the bond-energy configurations correspond to dimer configurations of the dual dice lattice in a similar manner, with one or three dimers touching each three-coordinated site of the dice lattice.

In these worm algorithms at $T=0$, the defect at the head of the worm corresponds to a monomer, {\em i.e.} a site on the dual lattice with no dimers touching it. The antidefect corresponds to an antimonomer on the same sublattice, {\em i.e.} a site with two dimers touching it. The initial defect-antidefect pair is created by simply picking a site at random and pivoting the dimer touching it to another unoccupied link. The antidefect is then propagated by pivoting successive dimers along a closed path, with probabilities chosen to preserve detailed balance. The updated dimer configuration of the dual lattice is then mapped back to a new spin configuration after the worm construction is complete. This flips an entire cluster of spins. At $T>0$, the worm construction is suitably generalized to work with more general defect anti-defect pairs.\cite{loopalgo} In this $T>0$ generalization\cite{loopalgo}, both the defect and the anti-defect correspond to dual sites with an {\em even} number of dimers touching them. These take the system out of the original configuration space of the equilibrium system, exactly as in the simpler $T=0$ construction.

The fact that all these worm constructions preserve detailed balance in a larger configuration space with one defect-antidefect pair allows for an interesting and simple method to calculate the corresponding correlation functions: The equilibrium defect-antidefect correlation function is simply proportional to the histogram of the head-to-tail separations measured during the worm construction.\cite{Alet_Ikhlef_Jacobsen,Sandvik_Moessner,Dorneich_Troyer} In the quantum rotor case, and in the context of worldline and SSE methods for bosonic systems, this corresponds to the imaginary time single-particle Green function.\cite{Alet_Ikhlef_Jacobsen,Sandvik_Moessner,Dorneich_Troyer,Alet_Sorensen_1} As we detail below, in the example studied in our work here, this corresponds to the correlation
function between half-vortices (with vorticities $\pm 1/2$) in the argument $\phi$ of the complex three-sublattice order parameter of the spin system.\cite{loopalgo} 

Apart from measuring the defect-antidefect correlator during worm construction, one can also measure various statistical properties of the worms themselves; the simplest of these is the distribution of worm lengths. This is of interest because the Monte-Carlo autocorrelation properties of such worm algorithms depend on the number of variables updated
in a single worm construction, which in turn depends on the distribution of worm lengths. For instance, the fractal structure and scaling properties of worms defined within the high temperature expansion have been studied previously.\cite{Janke_Neuhaus_Schakel} Properties of spin clusters defined by other cluster algorithms\cite{Swendsen_Wang, Wolff} have been numerically studied in the case of the critical two-dimensional Ising model\cite{Janke_Schakel} and found to be in agreement with theoretical predictions.\cite{Stella_Vanderzande,Saleur_Duplantier,Stanley,Vanderzande_Stella,Duplantier} Following the generalization of cluster algorithms to the fully frustrated square lattice,\cite{Kandel_Ben-Av_Domany_PRL} the properties of such clusters have also been studied extensively in that setting.\cite{Franzese} Since closed worms on the dual lattice define a cluster on the original lattice, properties of these clusters are also interesting from this point of view.  Statistics of worms constructed by a direct worm algorithm for a three dimensional spin ice model have also been studied, but less information seems to be available on worms in the corresponding two dimensional model.\cite{Jaubert_Haque_Moessner}

Part of the motivation for the present study  is our earlier observation that the autocorrelation properties of two rather different dual worm algorithms (the Deposition-Evaporation-Pivoting or DEP algorithm and the myopic algorithm)\cite{loopalgo} are determined by the universal exponent of the equilibrium spin-spin correlation function in the power-law three-sublattice ordered phase of frustrated Ising models on two  different two-dimensional lattices (triangular and kagome) over a range of temperatures. Since the properties of individual worms are expected to control the manner in which successive configurations decorrelate with each other,
we attempt to understand this universality by focusing here on a detailed study of the random geometry of these worms, specifically the number of steps taken to complete one worm, the number of distinct dual links flipped by one worm, and the number of Ising spins flipped by one worm.

Viewing each step of the worm construction as a position increment (step) of a random walker, the distribution of worm lengths is equivalent to the return time distribution of this random walker. In the power-law three-sublattice ordered phase of the equilibrium system, we find that this distribution also takes on a power-law form.  This motivates our detailed computational study of the persistence exponent $\theta$ and the dynamical exponent (fractal dimension) $z$ of this walk.
We provide convincing numerical evidence that the
  persistence exponent $\theta$ and the dynamical exponent $z$ of this random walk depend only on the universal power-law exponents of the underlying critical phase, and not on the details of the worm algorithm or microscopic Hamiltonian.  

Further, we argue that two key properties of this random walk follow directly from the detailed balance requirement obeyed by the worms, and the power-law correlations of the underlying equilibrium system: First, the steps of the walk are expected to be power-law correlated in time.  Second, the position distribution of the walker relative to its starting point is given by the equilibrium position distribution of a particle in an attractive logarithmic central potential of strength $\eta_m$, where $\eta_m$ is the universal power-law exponent of the equilibrium defect-antidefect correlation function of the underlying system. This latter observation is a particularly useful reformulation of the well-known idea that the head-to-tail separations measured during the worm construction directly measures the equilibrium defect-antidefect correlation function.\cite{Alet_Ikhlef_Jacobsen,Sandvik_Moessner,Dorneich_Troyer}. 

Using this reformulation, we derive a scaling relation,
\begin{equation}
  \label{scalinglaw}
  z=\frac{2-\eta_m}{1-\theta}\; ,
\end{equation}
that allows us to relate
  the dynamical exponent $z(\eta_m)$ of this process to its persistence exponent $\theta(\eta_m)$.

If this random walker had no memory (corresponding to the usual case of a Markovian random walker with dynamical exponent $z=2$) and was attracted to its starting point by a {\em static} logaritmic potential of strength $\eta_m$, the equilibrium position distribution (relative to its starting point) of this random walker would be consistent with the expected form of the distribution of head-to-tail displacements for our worms. Standard results\cite{Bray,Redner} on such random walks in a central logarithmic potential would immediately imply $\theta(\eta_m) = \eta_m/2$. Of course, this result is consistent with the more general scaling relation Eq.~\ref{scalinglaw}, since $z=2$ for the Markovian random walker.   

However, our results on the random geometry of the worms are not consistent with this simple picture of a Markovian random walker in a static potential. This is not entirely unexpected: As we detail below, this simple picture corresponds to an alternate dynamics in which the ``background'' dimer model has rapid thermal fluctuations allowing it to equilibrate very rapidly for each new position of the anti-defect, while the motion of the anti-defect pair is very slow in contrast. This may be viewed as a kind of ``Born-Oppenheimer limit'' of the dynamics. However, the actual worm dynamics being studied is as far from this Born-Oppenheimer limit as can be, since the background dimer configuration {\em only} changes each time the anti-defect moves.

Indeed, our computational study yields values of $z(\eta_m)$ and $\theta(\eta_m)$ that are consistent with the scaling relation Eq.~\ref{scalinglaw}  over a large range of values of the universal equilibrium exponent $\eta_m$. Throughout this range, the dynamical exponent is subdiffusive ($z>2$) and $\theta(\eta_m)$ is consistently larger than $\eta_m/2$. Thus we demonstrate that the worms constitute a discrete-time realization of a fractional Brownian motion characterized by these properties.

The rest of this paper is organized as follows: In Section.~\ref{Models} we provide a brief review of the models which are simulated by the worm algorithms studied here, along with a schematic description of their phase diagram. In Section.~\ref{Algorithms} we provide a quick summary of the DEP and myopic worm algorithms of Ref.~[\onlinecite{loopalgo}] whose properties we study. In Section.~\ref{Random_walker} we present our derivation of the scaling relation, Eq.~\ref{scalinglaw}, review the relevant facts about a Markovian random walker in a static central logarithmic potential and discuss how this simple Markovian walker scenario corresponds to the Born-Oppenheimer limit mentioned earlier, and sketch the connnection between the random geometry of our worms and the ensemble of overlap loops obtained by superposing two dimer configurations drawn independently from the equilibrium ensemble.  In Section.~\ref{Observables} we provide precise definitions of various properties of
the worm which are measured during the worm construction. In Section.~\ref{Results}, we summarize our results for these statistical properties of the worms including the persistence exponent $\theta$ and the dynamical exponent $z$. Finally
in Section.~\ref{Outlook}, we discuss some promising directions for future work.

\section{Models}
\label{Models}

\begin{figure}
	\includegraphics[width=\columnwidth]{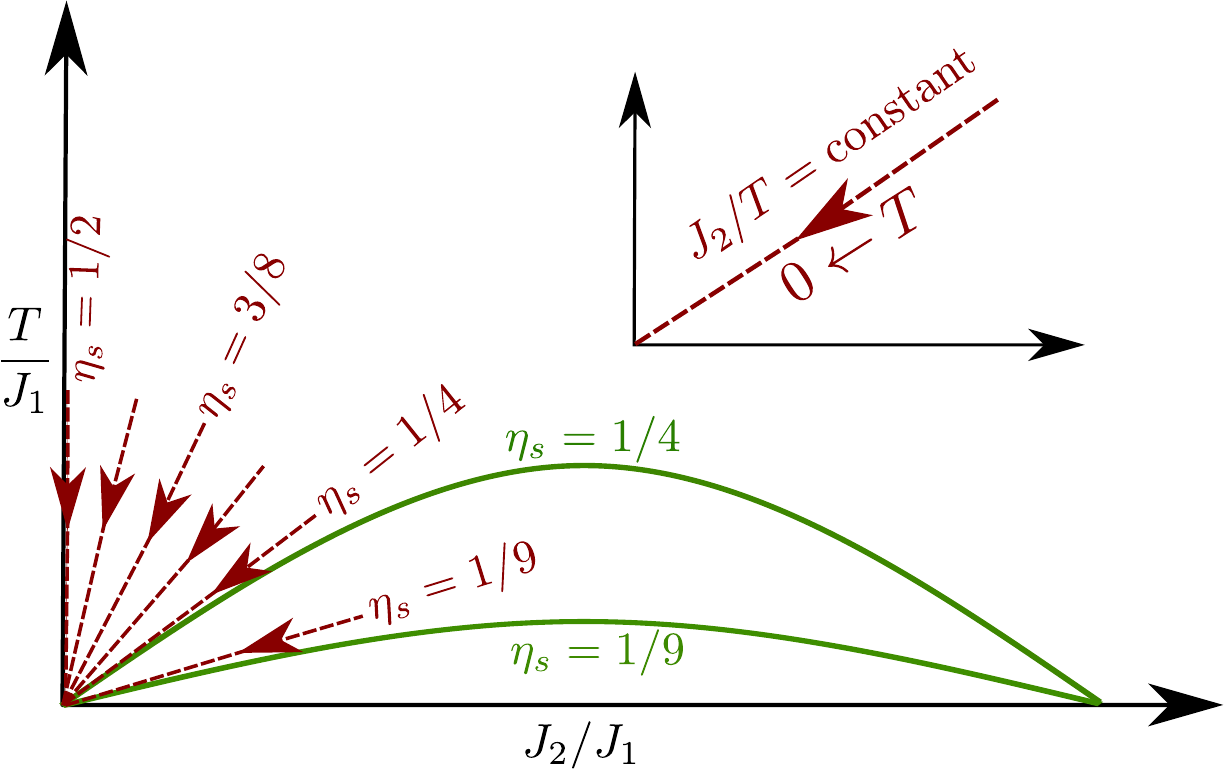} 
	\caption{\label{phase_diagram} The $T-J_2$ phase diagram of a triangular lattice antiferromagnet at a fixed nearest-neighbour antiferromagnetic coupling $J_1>0$. $J_2>0$ represents a ferromagnetic second neighbour interaction on the triangular lattice (thus, the sign conventions for $J_1$ and $J_2$ are opposite of each other). The dashed red lines represent the $T \rightarrow 0$ limit at a fixed value of $J_2/T$. In this limit, the critical exponent $\eta_s$ of the two-point spin correlation function at the three-sublattice wavevector goes from $1/2$ at $J_2/T = 0$ limit to $1/9$ as $J_2/T$ is monotonically increased to reach the threshold of long-range three-sublattice order. The solid green lines represent the lower and upper phase boundaries of the power-law three-sublattice ordered phase associated with the two-step melting of three-sublattice order at nonzero $J_2>0$. This critical phase has a continuously varying $\eta_s \in [1/9,1/4]$, with the lower (upper) limit being achieved at the lower (upper) phase boundary of this phase. See Sec.~\ref{Models} for a detailed discussion.
	}
\end{figure}

Ising models on triangular and kagome lattices with antiferromagnetic nearest neighbor interactions are among the simplest models of geometric frustration.\cite{Wannier,Kano_Naya}
For these models, the pattern of nearest-neighbour bond energies can be represented in terms of dimer models on the corresponding dual lattice (honeycomb and dice respectively).\cite{loopalgo} When further neighbour ordering interactions are absent, there is a macroscopic degeneracy of minimum energy spin configurations, which corresponds to a $T=0$ ensemble of dimer configurations on the dual lattice.
For the triangular lattice antiferromagnet, this $T=0$ ensemble is made up of all
perfect matchings (fully-packed dimer configurations) on the honeycomb lattice, while the $T=0$ dimer configurations on the dice lattice have exactly one dimer touching each three-coordinated site and an even number of dimers touching each six-coordinated site. The former ensemble has power-law dimer correlations with power-law exponent $\eta_d=2$ (at the uniform and the three-sublattice wavevectors). This corresponds to power-law correlations for the spins at the three-sublattice wavevector, with power-law exponent $\eta_s=1/2$ at $T=0$.\cite{Wannier,Stephenson} The kagome lattice antiferromagnet in this limit is a short-range correlated spin liquid,\cite{Kano_Naya} corresponding to short-range dimer correlations.

At $T=0$ for the nearest neighbour triangular antiferromagnet, the relationship between $\eta_d$ and $\eta_s$ can be understood in terms of a coarse-grained height model\cite{Alet_Ikhlef_Jacobsen,Blote_Hilborst,Nienhuis_Hilhorst_Blote,Fradkin_etal} for the ensemble of fully-packed dimer configurations on the honeycomb lattice. In this representation, the spin operator at the three-sublattice wavevector corresponds to $\exp(i \pi h)$ (where $h$ is the height field)
while the dimer operator has a uniform part given in terms of the gradient $\nabla h$ and a second piece $\exp(2\pi i h)$ at the three-sublattice wavevector. Thus, one can think of this height field $h$ as being proportional to the phase $\phi$ of the local three-sublattice order parameter of the spin system: $h \equiv \phi/\pi$. 

This coarse-grained effective field theory is useful because the action for $h$ (equivalently for $\phi$) is Gaussian, characterized by a single dimensionless stiffness $g$. Dimer correlations at the
uniform wavevector fall of as $1/r^2$ independent of the stiffness of the height model, while correlations at the three-sublattice wavevector decay with power-law exponent $\eta_d(g)$ controlled by the stiffness of the height model. Spin correlations at the three-sublattice wavevector fall of as a power law with exponent $\eta_s$ (with $\eta_d = 4\eta_s$). When all fully-packed dimer configurations have equal weight (as is the case for the nearest neighbour antiferromagnet in the $T \to 0$ limit),  $\eta_d=2$ and $\eta_s=1/2$.

A second-neighbour ferromagnetic interaction $J_2$ on the triangular lattice, with $J_2 \propto T$ in the $T \rightarrow 0$ limit, is equivalent to an attractive interaction favouring columnar three-sublattice order in the fully-packed dimer model that describes this limit. This interaction gives rise
to a $\eta_d <2$ and $\eta_s < 1/2$ as shown in Fig.~\ref{phase_diagram}. Indeed, $\eta_s$ decreases monotonically with increasing $J_2/T$ (in this zero temperature limit), until the system
develops long-range three-sublattice order when $\eta_s =1/9$ is reached.\cite{Nienhuis_Hilhorst_Blote} In the coarse-grained height representation, this is understood by noting that $J_2/T$ tunes the stiffness $g$ of the height model, thereby influencing the value of $\eta_s$ (and of $\eta_d=4\eta_s$). When $\eta_s = 1/9$, the leading allowed cosine interaction $\cos(6\pi h)$ becomes relevant, driving the transition to three-sublattice order.

Monomers in this fully-packed dimer model correspond, in the Coulomb gas (CG) description of the coarse-grained height model,\cite{Alet_Ikhlef_Jacobsen,Fradkin_etal} to a magnetic charge $\pm 1$ on $A/B$ sites of the honeycomb lattice (likewise, antimonomers have magnetic charge $\mp 1$ on $A/B$ sites of the honeycomb lattice). As a result, the monomer-antimonomer correlator decays as a power law with an exponent $\eta_m=1/\eta_d = 1/4\eta_s$. In terms of the argument $\phi$ of the complex three-sublattice
order parameter of the spin model, these monomers and antimonomers are half-vortices since $\phi \equiv \pi h$ and each monomer or antimonomer results in a height ambiguity of $\Delta h = \pm 1$ along any path that encircles the defect once.

At nonzero $T$, the dimer representation of bond energies now allows  three-coordinated sites touched by three dimers or one dimer, greatly increasing the entropy of allowed configurations. The sites touched by three dimers correspond to {\em vortices} in the phase $\phi$ of the three-sublattice order parameter of the spin model. The worm algorithm now makes other defects (Section.~\ref{Algorithms}) apart from monomers and antimonomers; like monomers and antimonomers, these too take the system outside the physical subspace of the spin model during the construction of the worm.  These defects can again be thought of as half-vortices in the argument $\phi$ \cite{Kedar_PRL,Chern_Tchernyshyov,Jose_etal} of the Fourier component of the spin density at the three-sublattice wavevector. 

A fixed nonzero value of second-neighbor ferromagnetic interaction induces long-range three-sublattice order on both lattices at low enough temperature.  This melts via a two-step process, wherein the intermediate state has power-law  spin correlations at the three-sublattice wavevector, with power-law exponent $\eta_s$  that ranges from $1/9$ (at the low-temperature end) to $1/4$ (at the high-temperature end) as shown in Fig.~\ref{phase_diagram}.\cite{Nienhuis_Hilhorst_Blote,Jose_etal,Landau,Wolf_Schotte} As noted earlier, the low-temperature boundary of the power-law ordered state corresponds to the threshold at which the leading cosine interaction $\cos(6\pi h)$ becomes relevant. Conversely, the high-temperature boundary corresponds to the threshold at which vortices in the three-sublattice order parameter of the spin model become relevant. Finally, we note that when spin correlations display power-law three-sublattice order, the dimer correlations also have a power-law form, with exponent $\eta_d=4 \eta_s$.

Since the power-law phase is described by
a Gaussian theory for $\phi$ characterized again by a single dimensionless stiffness $g$, the defect-antidefect correlator is again expected to decay with exponent $\eta_m=1/4\eta_s=1/\eta_d$ (where
$\eta_d$, the dimer correlation exponent, is again related to the power-law exponent $\eta_s$ via $\eta_d = 4 \eta_s$). The values $\eta_m <1$ can only be accessed in the zero temperature limit on the triangular lattice by tuning $J_2/T$, while the power-law three-sublattice ordered phase at nonzero temperature corresponds to the range $\left((1/2,9/4 \right)$) for $\eta_m$ on both lattices.

With this background, we use the previously developed DEP and myopic algorithms\cite{loopalgo}  in the triangular case and the myopic
algorithm in the kagome case to simulate the classical Ising model
\begin{eqnarray}
H &=& J_1\sum_{\langle RR' \rangle} \sigma_R \sigma_{R'} - J_2 \sum_{\langle \langle RR' \rangle \rangle} \sigma_R \sigma_{R'}  \;  , \nonumber \\
&&
\end{eqnarray}
where $\langle R R' \rangle$ and $\langle \langle R R' \rangle \rangle$ denote nearest-neighbor and next-nearest-neighbor links of the lattice in question, and $\sigma_R = \pm 1$
are the Ising spins on sites $R$ of the triangular or kagome lattice. In our convention, $J_{1} >0$ ($J_2 > 0$) corresponds
to an antiferromagnetic (ferromagnetic) coupling. We focus here on the case with $J_1>0$ and $J_2>0$, and study the statistics of worms generated by these algorithms in the power-law three-sublattice ordered phase on both lattices.

\section{Algorithms}
\label{Algorithms}

In this section we provide a brief description of the algorithms developed in Ref.~\onlinecite{loopalgo}, whose statistics we wish to study here.

\begin{figure*}
	\includegraphics[width=\textwidth]{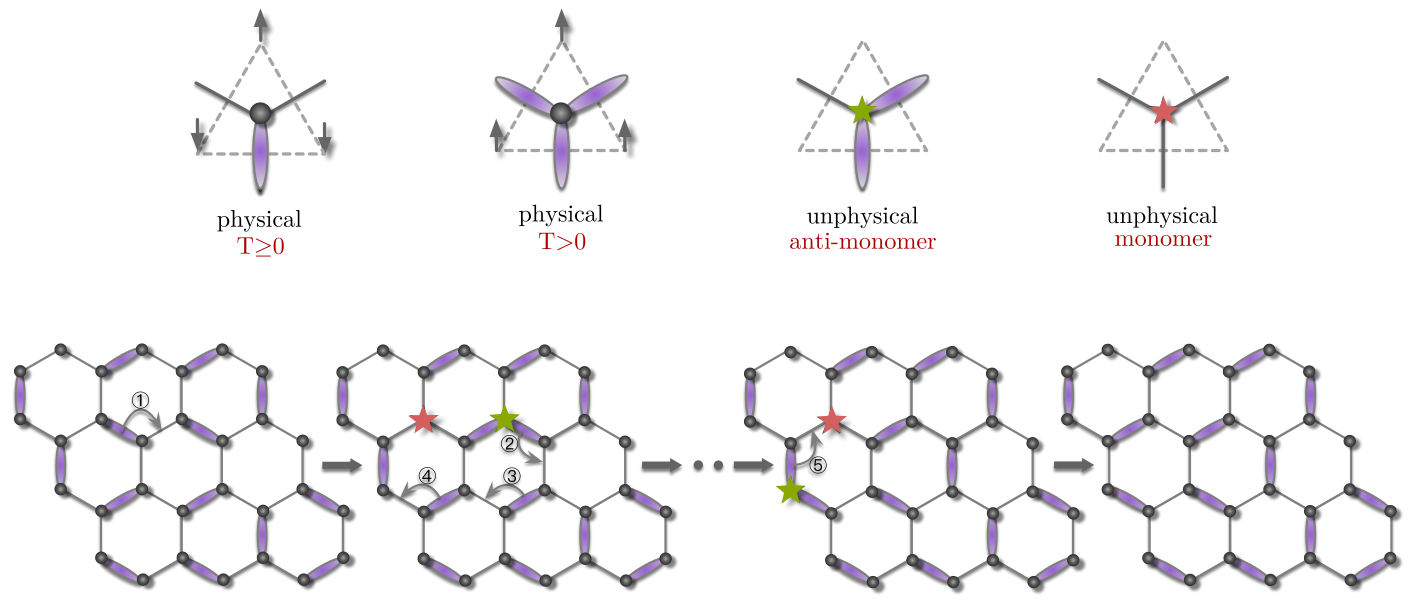} 
	\caption{\label{worm} 
		The first row shows the allowed dimer configurations and defect/anti-defect configurations of the three-coordinated sites of the honeycomb lattice (solid lines) dual to the triangular lattice (dashed lines). Dimers occupy bonds on the dual lattice and intersect a frustrated bond on the triangular lattice. For Ising spins with antiferromagnetic nearest-neighbour exchange couplings on the triangular lattice, a frustrated bond connects Ising spins that point in the same direction. Defect/antidefect dimer configurations do not represent physical spin configurations, and hence take the dimer model out of the physical configuration space of the Ising antiferromagnet. The second row shows a worm  propagating on the dual honeycomb lattice. Consecutive steps of the worm algorithm are numbered. Step $1$ of the worm algorithm (in both the DEP and the myopic algorithms) involves pivoting a dimer to create a pair of defect-antidefect sites (marked by five-pointed stars in the figure) on the dual lattice. As the worm propagates, the antidefect moves with the head of the worm, whereas the defect, associated with the tail of the worm, remains static at the starting site. The worm construction ends when the head reaches the tail again, causing the defect and antidefect to annihilate to produce a new physical dimer configuration that can be mapped to a legitimate configuration of the Ising spins so long as some parity constraints on the winding numbers are preserved.  See Sec.~\ref{Algorithms} for a detailed discussion.
	}
\end{figure*}

As mentioned in the introduction, these worm algorithms are defined on the dual lattice, and work with the dimer representation of the frustrated Ising antiferromagnet on the triangular and kagome lattices. At nonzero temperatures on the triangular lattice, each triangle has either one or three frustrated bonds. This translates to either one or three dimers touching each lattice site of the dual honeycomb lattice. Similarly on the kagome lattice at nonzero temperatures, every spin configuration corresponds to a dimer configuration in which each three-coordinated site of the dual lattice is touched by either one or three dimers, while each six-coordinated site of the dual lattice is touched by an even number of dimers.  Thus, at nonzero temperatures, the configuration space of the dual dimer model is larger than that of the usual fully-packed dimer model. The dimer configurations corresponding to various spin configurations have been shown in Fig.~\ref{worm}.

As noted earlier, a power-law three-sublattice ordered phase is obtained on  the triangular lattice both at $T = 0$ (with a ferromagnetic $J_2 \propto T$) and at $T > 0$ (associated with the melting of long-range three-sublattice order). On the kagome lattice, power-law three-sublattice order is obtained only at $T > 0$ (associated with the melting of long-range three-sublattice order). In Ref.~\onlinecite{loopalgo}, two worm algorithms were introduced for the triangular lattice model (the DEP and myopic algorithms), only one of which (the myopic algorithm) generalizes to the kagome lattice.\cite{loopalgo} Thus, we have five different settings in in which we can test the idea that the statistics of worms in the power-law three-sublattice ordered phase depends only on the exponent $\eta_s$, independent of the details
of the worm construction and form of Hamiltonian: One can study
the worm statistics of both algorithms in the $T=0$ power-law phase as
well as the $T>0$ power-law phase on the triangular lattice, and one can
also study the worm statistics of the myopic algorithm in the $T>0$ power-law phase on the kagome lattice. [Both the DEP and myopic worm algorithms, though developed for the larger dual configuration space at $T>0$, reduce in an obvious way at $T=0$ to previously known implementations of worm algorithms for dimer models on the honeycomb and dice lattices\cite{loopalgo}].

Both algorithms proceed by first translating the spin configuration into a dual dimer configuration, with each dimer configuration thus assigned a Boltzmann weight of the parent spin configuration. Next, we update the dimer configuration using these worm algorithms. The DEP and myopic worm algorithms differ in the way they perform this
update. While the DEP worm algorithm keeps track of the local dimer environment near the head of the worm at every step of the worm construction, the myopic worm algorithm does not keep track of the local dimer environment near the
head of the worm at alternate steps (it is thus ``myopic'' or short-sighted at alternate steps).
The details of the worm construction protocol for both algorithms, and the proofs
that these protocols obey detailed balance, have already been discussed extensively
in Ref.~\onlinecite{loopalgo}. Here, we confine ourselves to providing a simple example of the worm construction in the $T \to 0$ limit in Fig.~\ref{worm}.

Since detailed balance is explicitly satisfied, the dimer configuration obtained after the worm is constructed can always be accepted. However, there is one subtlety when it comes to accepting the corresponding spin configuration: We work with periodic boundary conditions along $\hat{x}$ and $\hat{y}$ of the triangular and kagome lattices. This translates to constraints on the parity of the global winding number of the corresponding dimer model (For details on preserving detailed balance and winding number constraints see Ref.~\onlinecite{loopalgo}). Thus, after the worm construction, only updated dimer configurations which satisfy this constraint can be translated back to the spin configuration. Therefore, one has to occasionally reject a worm which winds around the lattice, if the result leads to a dimer configuration in an illegal winding number sector.

The worm algorithm typically propagates the head of the worm along a complicated path on the dual lattice. As we discuss in more detail in the next section, this path can have frequent self-intersections in addition to backtracking. As a result, the correspondence between the geometry of this worm and the degrees of freedom it updates is not entirely straightforward. Nevertheless, it is straightforward to construct a new spin configuration from the new dimer configuration obtained after the head of the worm has recombined with the static tail (if the new dimer configuration does not obey the winding number parity constraints necessary for being in correspondence with a spin configuration, we do not keep track of this worm). Since each valid dimer configuration corresponds to two spin configurations related by a global spin flip, we randomly chose one of these when translating back to the spin configuration.

\section{Random walk considerations}
\label{Random_walker}
\subsection{Scaling relation between $\theta$ and $z$}
Our starting point
is the well-known observation,\cite{Alet_Ikhlef_Jacobsen,Sandvik_Moessner,Dorneich_Troyer} alluded to in the introduction, that the histogram of head-to-tail distances of the worm is given by the equilibrium defect-antidefect correlator $C_m(\vec{r})$. Viewing each step of the worm construction as a position increment (step) of a random walker, this translates to the requirement that the walk has a long-time steady state distribution of position given by $C_m (\vec{r})$. In other words, the histogram of positions $\vec{r}$, accumulated during the walk, must be proportional to $C_m (\vec{r})$.
If we choose a normalization convention whereby this histogram measures the ratio of the number of times the head to tail separation is $\vec{r}$ to the number of returns to the origin, {\em i.e.} the number of times the head to tail separation is $\vec{r}$ {\em per worm},  then
the mean return time is given as  $\langle \tau_r \rangle = v$, with
$v \equiv \sum_{\vec{r}} C_m(\vec{r})$, where
the sum extends over $L^2$ sites of the finite lattice. If this sum is dominated by
contributions near the upper cutoff in distance, we expect $v  \sim   L^{2-\eta_m}$. This is
true for all $\eta_m < 2$.

Next we note that the average return time can also be expressed in terms of the probability distribution $P(\tau_r) $ of first return time $\tau_r$ of this walker by writing $\langle \tau_r \rangle = \sum_{\tau_r} \tau_r P(\tau_r)$.
Assuming that the power-law form $P(\tau_r) \sim 1/\tau_r^{\theta+1}$ persists up
to a system-size dependent cutoff scale  $\tau_{\textrm{cutoff}}(L) \sim L^z$, where
$z$ is the dynamical exponent for the random walk, we obtain $\langle \tau_r \rangle \sim L^{z(1-\theta)}$ whenever the sum is dominated by the contributions near the upper cutoff.
This is true for all $\theta < 1$.

Comparing these two predictions for the $L$ dependence of the mean return time, we arrive at the scaling relation of Eq.~\ref{scalinglaw} which is valid when $\eta_m < 2$ and $\theta < 1$.
Thus, the dynamical exponent $z$ is in general not fixed to the usual Markovian
random walk value of $z=2$. This appearance of a non-Markovian value of $z$
in our description of the worms should be interpreted in the following way: The underlying worm algorithm is Markovian. The probability table that guides the choice of the next
step in the worm construction depends only on the current configuration of the system.
However, when one only focuses on the position of the head relative to the fixed
tail of the worm, one is tracing out all information about the rest of the system, {\em i.e.} losing information about the power-law correlated background dimer liquid. This
projected process is non-Markovian, in the sense that it depends in principle
on the entire history of previous positions of the head, and may be expected to have a memory that falls off as a power-law in time since the equilibrium power-law correlations of the spin and bond energy variables also give rise to correlations between the successive steps taken by the head of the worm. It is this effectively non-Markovian dynamics that
is being described in terms of a $z$ different from $z=2$.

Note that if $z$ had been fixed at $z=2$, as is the case for the usual Markovian random walk, our scaling relation could have been used to predict (incorrectly) that $\theta(\eta_m) = \eta_m/2$. This connection between the ansatz $\theta(\eta_m) = \eta_m/2$ and the Markovian value $z=2$ is amenable to a simple interpretation via a toy model of an ordinary random walker in an attractive logarithmic potential of strength $\eta_m$. 

In the next section, we outline a different dynamics for which this toy model of an ordinary random walker in a central logarithmic potential is expected to yield exact results. By contrasting the actual worm dynamics with this alternate dynamics, we isolate and pinpoint the precise features of the worm dynamics responsible for the interesting non-Markovian behaviour observed by us.

\subsection{Alternate dynamics: The ``Born-Oppenheimer limit''}
Let us imagine that the underlying dimer system has other ways of equilibrating, and does not rely entirely on the motion of the worms to reach equilibrium. Indeed, let us consider an alternate dynamics, in which the dimer system rapidly reaches equilibrium at each new position of the antidefect during the motion of the worm (recall that the defect is held fixed at its initial location and the worm construction involves the motion of the antidefect). For concreteness, one can imagine this is achieved by performing a very large number, $N_{\rm local}$, of local ring-exchange moves that leave the defect and the antidefect unchanged but allow the dimer configuration to equilibrate for these fixed locations of the defect and antidefect. Thus, in this alternate dynamics, the time scale for equilibrating the dimer system is much shorter than the time scale over which the worm head moves. Motivated by the analogy to ``slow nuclei'' and ``fast electrons'' in  the Born-Oppenheimer theory of molecular dynamics, we dub this $N_{\rm local} \to \infty$ limit the Born-Oppenheimer limit. 

Defining a coordinate system in which the static defect is at the origin, the antidefect in this limit will see a static central potential of entropic origin, arising from the dependence of the equilibrium dimer partition function on the location of the antidefect. In other words, we may now legitimately view the power-law form of the equilibrium defect-antidefect correlation function as being the result of a static attractive logarithmic interaction  $V(\vec{r}) \equiv -\ln(C_m (\vec{r})) = \eta_m \ln(r)$ between the head and the tail (recall that the tail is held fixed while the head is mobile in this alternate dynamics, exactly as in the original worm dynamics). 
Thus, in this Born-Oppenheimer limit, which we view as a simple toy model for the motion of the worm, the head executes an ordinary Markovian random walk in an attractive central logarithmic potential. This walk starts at a site adjacent to the origin and ends when it returns to the origin for the first time. The worm
length in this picture is mapped to the time  $\tau_r$  of first return to origin of this random walk. As a result, the worm length distribution is given by the probability distribution $P(\tau_r)$ of return times, and is expected to have a power-law form $P(\tau_r) \sim 1/\tau_r^{\theta+1}$, where $\theta$ is the persistence exponent.

The Fokker Planck equation for a Brownian walker in an attractive logarithmic central potential in dimension $d$ can be transformed\cite{Bray,Redner}
using radial coordinates to the equation for a free Brownian walker in an effective dimension $d'$, with
\begin{equation}
d'=d-\eta_m
\end{equation}
The probability distribution of the first return time $\tau_r$ of a Markovian random walker in dimension $d'$ can be obtained using the corresponding Green function with absorbing
boundary conditions at the origin. This predicts the large-$\tau_r$ form:
\begin{equation}
	P(\tau_r) \sim 
	\begin{cases}
		1/\tau_r^{2-(d'/2)}, & \text{for }  d' < 2 \\
		1/(\tau_r\ln^2(\tau_r)), & \text{for } d' = 2 \\
		1/\tau_r^{(d'/2)}, & \text{for } d' > 2 \; .
	\end{cases}
\end{equation}

Thus, for a Markovian random walker
\begin{equation}
\label{Theta}
\theta =
\begin{cases}
1-(d'/2), & \text{for }  d' < 2 \\
(d'/2)-1, & \text{for }  d' > 2 \; .
\end{cases}
\end{equation}

Since $d=2$ in our case and $\eta_m \in \left(1/2,9/4 \right)$, $d' \equiv 2-\eta_m$ is always less than $2$. Thus, when $z=2$, these standard results for a Brownian walker give $\theta = \eta_m/2$, valid when  $\eta_m \in (1/2, 2)$. This is consistent with the more general scaling relation between $z$ and $\theta$. For $\eta_m \geq 2$, {\em i.e.} close to the ordering transition at which the power-law spin order gives way to long range order, $d'$ turns negative and the analysis leading to Eq.~\ref{Theta} breaks down (in this regime, the more general scaling relation between $\theta$ and $z$ also breaks down). 

As emphasized earlier, the actual value of $\theta(\eta_m)$ obtained from our numerical work deviates significantly from this Markovian model. The numerical results consistently give a larger value of $\theta(\eta_m)$ over the entire range of $\eta_m$. This observed value of $\theta$ is accompanied by a measured dynamical exponent that is sub-diffusive ($z>2 $) in the entire range studied. Nevertheless, in the entire range studied, $\theta$ and $z$ satisfy the scaling relation derived in the previous subsection.

\subsection{Decomposition of a worm trace into overlap loops}
The trace of a worm, {\em i.e.} the trajectory of its head before it eventually returns to and recombines with the tail, is in general not a simple loop. Indeed, as will be clear from our numerical results in the next section, it can intersect itself often, in addition to backtracking along previously traversed segments. The question then arises: Can this complicated trace be related in a useful way to the changes in the dimer configuration produced as a result of this worm?

To answer this question, let us first consider the zero temperature limit on the triangular lattice. In this limit, the dual configuration space reduces to that of the hard-core dimer model on the honeycomb lattice. An elegant geometric characterization of the difference between the final dimer configuration and the initial dimer configuration is now possible in terms of the overlap loop diagram obtained by superposing these two dimer configurations. Since each site is covered by exactly one dimer in both the initial and final dimer configuration, each loop in this diagram is a simple loop. All these simple loops admit a consistent orientation: One constructs each of them by starting with any $A$ sublattice site on the loop and alternately following the dimers from the final and initial configurations until one returns to the starting point.

Since both the final and the initial dimer configurations are drawn from the equilibrium ensemble corresponding to the fully-packed dimer model (with interactions corresponding to the limiting $T \to 0$ value of $J_2/T$), long-wavelength properties of these dimer configurations admit a description in terms of the coarse-grained height model alluded to earlier. If the initial and final dimer configurations had been drawn completely {\em independently} from this equilibrium ensemble, so that one was in effect considering the overlap between two {\em independent copies} of the equilibrium system, it would be possible to characterize the large-scale properties of the overlap loop diagram in terms of the statistics of contour lines of equal height of a fluctuating height field with a Gaussian action (see for instance Ref.~\onlinecite{Desai_Pujari_Damle}, where such a characterization was used as a diagnostic for a novel bilayer Coulomb phase in a bilayer dimer model). 

Much is known about the statistics of contour lines of a fluctuating Gaussian height field. For instance, the probability that two points separated by $\vec{r}$ lie on the same contour falls off as $1/r^{2x_1}$; crucially, the exponent $2x_1=1$ is independent of the dimensionless stiffness $g$ of the Gaussian height action. This is in sharp contrast to the continuously varying exponent $\eta_m(g)$ which characterizes the probability that the head-to-tail displacement is $\vec{r}$ for our worm dynamics. It is also known that the persistence exponent $\tau-1$, which characterizes the power-law distribution of size of the contour loop that passes through a randomly chosen point, and the dynamical exponent $z_{\rm contour}$ that characterizes the fractal nature of the contour loops, are both independent of the stiffness $g$ of the fluctuating Gaussian height field:  $\tau-1 = 4/3$, and
$z_{\rm contour} \equiv D_f = 3/2$.\cite{Kondev_Henley} 

These values of $z_{\rm contour}$, $\tau -1$ and $2x_1$ obey a scaling relation\cite{Kondev_Henley}  $D_f = (2-2x_1)/(3-\tau)$ which follows from a scaling argument entirely analogous to the analysis that led us to our scaling relation Eq.~\ref{scalinglaw}. Recent work on the statistics of overlap loops obtained from dimer configurations of two layers of a bilayer system in a novel bilayer Coulomb phase has found excellent agreement with these values of $D_f$ and $\tau$; indeed these values were found to control the {\em full finite-size scaling function} of the size of these overlap loops, {\em independent} of the stiffness of the associated coarse-grained height field.\cite{Desai_Pujari_Damle}

How does the random geometry of worms studied here relate to this statistics of overlap loops? A partial answer comes from noting
that each self-intersection of the worm results in an overlap loop split off from the rest of the worm. Thus, the overlap loop diagram obtained by superposing the initial and final dimer configurations can be viewed as being the result of `resolving' each self-intersection of the worm to obtain a collection of simple ``daughter loops''. Clearly, the number of such daughter loops obtained from a completed worm is itself a random variable, as are the sizes of these daughter loops. The random geometry of our worms should therefore be viewed as a convolution of the properties of a variable number of these daughter loops. While we have not explored this connection in any detail, we mention it here since it provides a useful perspective on our results. In particular, since a single worm is a convolution of a variable number of overlap loops, and the detailed nature of this convolution depends continuously on the equilibrium exponent $\eta_m$ (equivalently on the stiffness $g$ of the equilibrium Gaussian height action), this connection provides a rationale for the fact that the worm-exponents $\theta(\eta_m)$ and $z(\eta_m)$ vary continuously with $\eta_m$, while the corresponding overlap-loop exponents $\tau-1$ and $D_f$ are independent of the stiffness $g$.

Finally, we note that this connection becomes somewhat less direct at a microscopic level in the power-law ordered phase at nonzero temperature. In this case, the dimer configurations have a nonzero density of dual sites touched by three dimers instead of one. Therefore, the superposition of two such dimer configurations does not admit a unique decomposition into set of non-overlapping simple loops. Therefore, it would be useful to perhaps explore alternate microscopic constructions that attempt to connect our worms to a convolution of some suitably defined simple loops that represent microscopic realizations of the contour lines of the Gaussian free field that controls the long-wavelength description of this nonzero temperature power-law phase. We have not attempted this here.

\begin{figure*}
	\includegraphics[width=\textwidth]{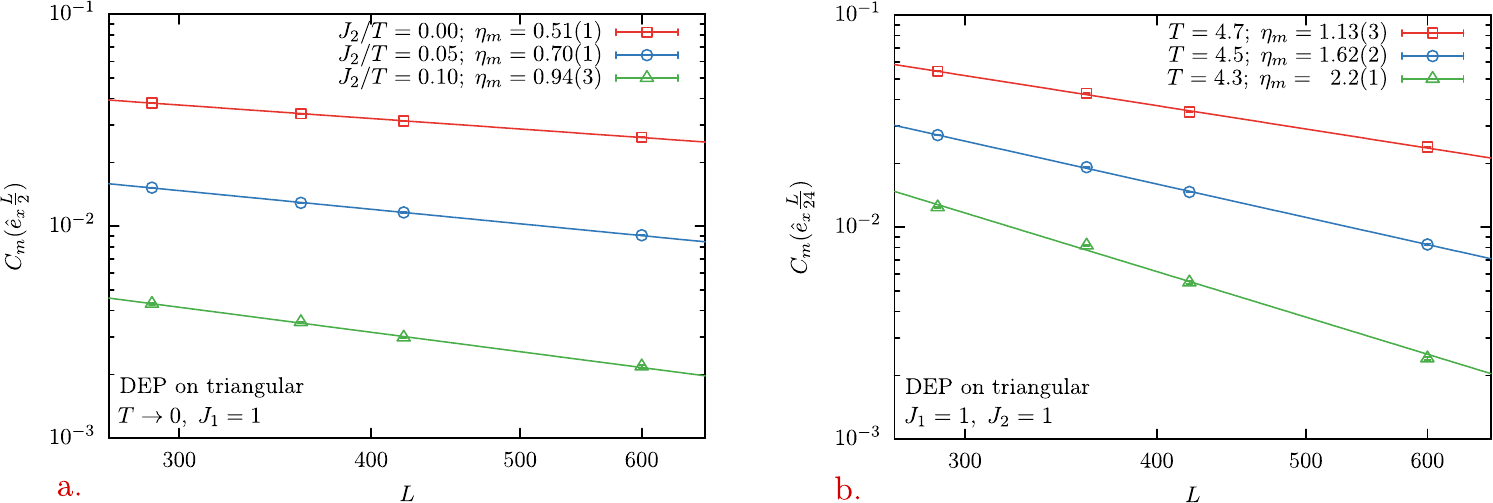}
	\caption{\label{histdefect}The lattice size $L$ dependence of the defect-antidefect correlation function $C_{m} \left( \hat{e_x} \frac{L}{a}  \right)$ at separation $\hat{e_x} \frac{L}{a}$ on a periodic $L \times L$ triangular lattice for the DEP worm algorithm for values of parameters at which the system has power-law three-sublattice order [\textcolor{Red}{a.}] in the $T \rightarrow 0$ limit, and [\textcolor{Red}{b.}] at a nonzero $T$. Line denotes fit to a power-law form $C_m\left( \hat{e_x} \frac{L}{a}  \right) \propto 1/L^{\eta_m}$. Note that any point in the $T>0$ power-law three-sublattice ordered phase has $\eta_m>1$, while smaller values of $\eta_m$ can be accessed in the $T=0$ limit considered here. Corresponding plots for the myopic worm algorithm on the triangular and kagome lattices are shown in Fig.~SM.1. of the Supplemental Material.\cite{suppmat}
	See Sec.~\ref{Observables} and Sec.~\ref{Results} for a detailed discussion.}
\end{figure*}
\begin{figure*}
	\includegraphics[width=\textwidth]{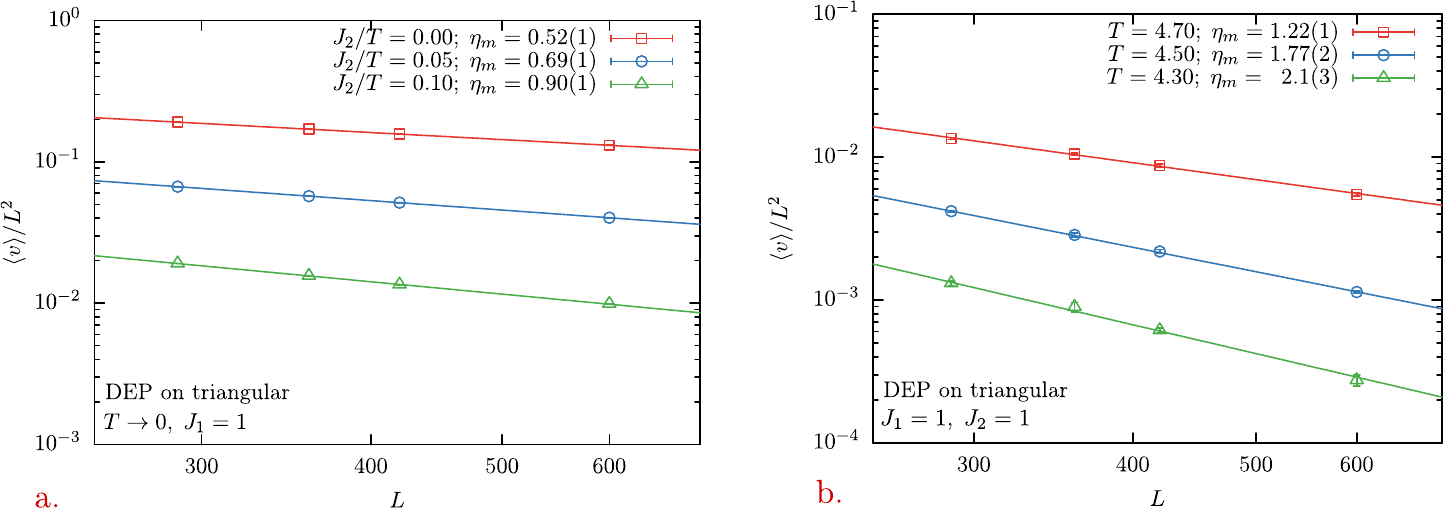}
	\caption{\label{mean_visits} The lattice size $L$ dependence of the average number of dual lattice sites visited per worm, $\avg{v}$, for the DEP worm algorithm on a periodic $L \times L$ triangular lattice for values of parameters at which the system has power-law three-sublattice order [\textcolor{Red}{a.}] in the $T \rightarrow 0$ limit, and [\textcolor{Red}{b.}] at a nonzero $T$. Since $\avg{v}/L^2 \sim 1/L^{\eta_m}$, the power-law fits give us an alternate measurement of $\eta_m$. Corresponding plots for the myopic algorithm  are shown in Fig.~SM.2 of the Supplemental Material.\cite{suppmat} See Sec.~\ref{Observables} and Sec.~\ref{Results} for a detailed discussion.
	}
\end{figure*}
\begin{figure}
	\includegraphics[width=\columnwidth]{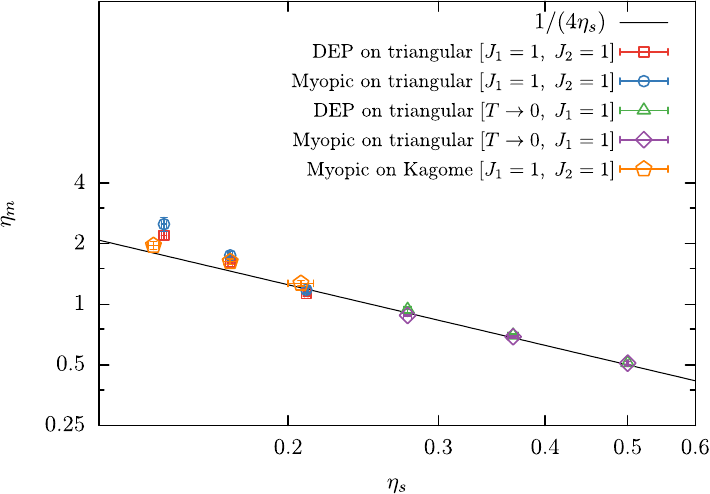}
	\caption{\label{eta_defect}$\eta_m$ extracted from $L$ dependence of the defect-antidefect correlators (Fig.~\ref{histdefect}) plotted as a function of $\eta_s$, the exponent of the power-law spin-spin correlations at the three-sublattice wavevector. The line denotes the theoretically expected dependence $\eta_m^{\textrm{predicted}} =1/4\eta_s$. See Sec.~\ref{Observables} and Sec.~\ref{Results} for a detailed discussion.
	}
\end{figure}

\begin{figure*}
	\includegraphics[width=\textwidth]{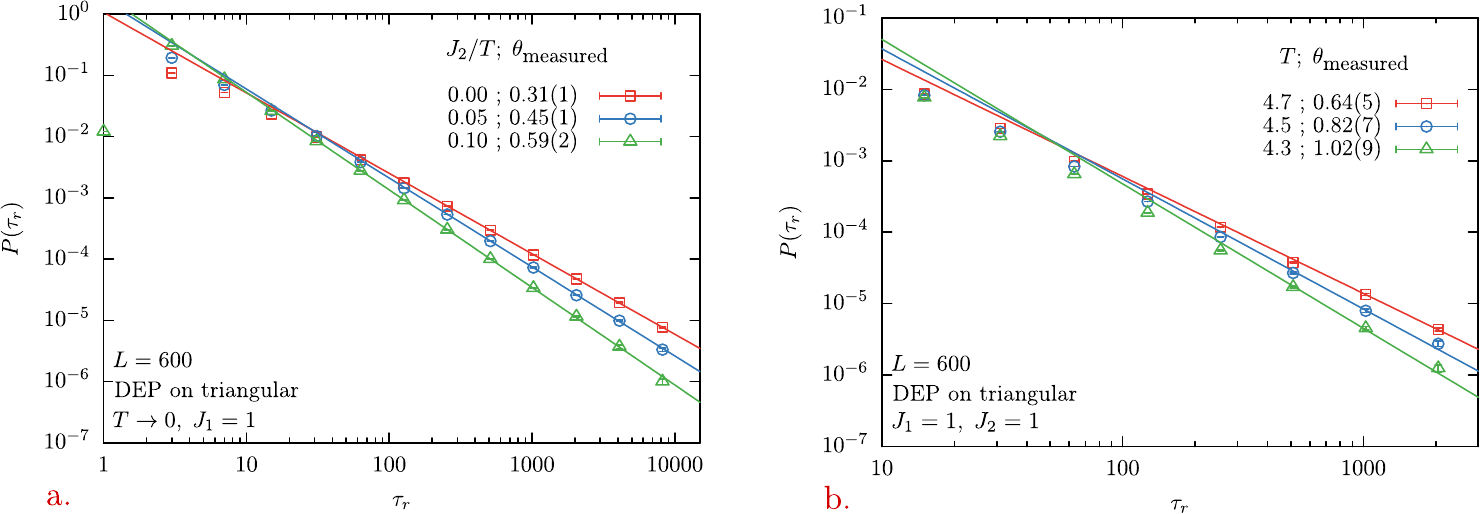}
	\caption{\label{histvisited} Probability distribution $P(\tau_r)$ of the worm length $\tau_r$, {\em i.e.} the number of sites $\tau_r$ of the dual lattice visited by a worm of the DEP worm algorithm on a periodic $L \times L$ triangular lattice for values of parameters at which the system has power-law three-sublattice order [\textcolor{Red}{a.}] in the $T \rightarrow 0$ limit, and [\textcolor{Red}{b.}] at a nonzero $T$. Lines denote fits to a power-law form $P(\tau_r)\propto 1/{\tau_r}^{1+\theta_{\text{measured}}}$. Corresponding plots for the myopic worm algorithm on the triangular and kagome lattice are shown in Fig.~SM.2 of the Supplemental Material.\cite{suppmat} Note that the value of $\theta_{\text{measured}}$ for $T=4.3$ is at the edge of validity of the scaling relation of Eq.~\ref{scalinglaw}. See Sec.~\ref{Observables} and Sec.~\ref{Results} for a detailed discussion.
	}
\end{figure*}
\begin{figure}
	\includegraphics[width=\columnwidth]{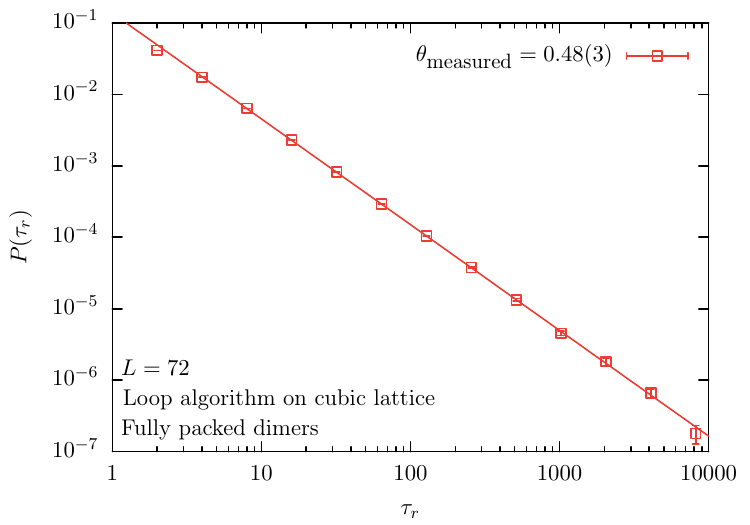}
	\caption{\label{histvisited-dimers} Probability distribution $P(\tau_r)$ of the worm length $\tau_r$, {\em i.e.} the number of sites $\tau_r$ on a periodic $L \times L \times L$ cubic lattice which are visited by a non winding worm of the dimer worm algorithm for the fully packed dimer model on a $L=72$ cubic lattice. Line denotes fit to a power-law form $P(\tau_r)\propto 1/{\tau_r}^{1+\theta_{\text{measured}}}$. See Sec.~\ref{Observables} and Sec.~\ref{Results} for a detailed discussion.
	}
\end{figure}

\begin{figure*}
	\includegraphics[width=\textwidth]{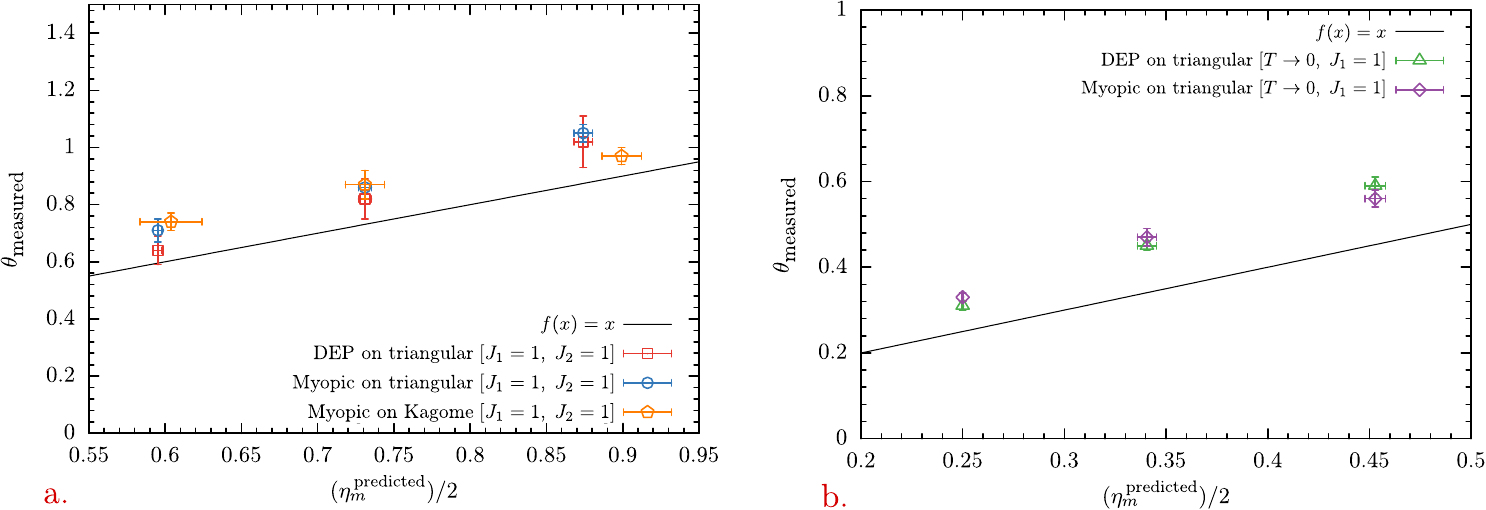}
	\caption{\label{theta} The persistence exponent $\theta$ (extracted from fits to $P(\tau_r)$) displayed as a function of $(\eta_m^{\textrm{predicted}})/2 \equiv 1/(8 \eta_s)$ in simulations employing the DEP and myopic worm algorithms for values of parameters at which the system has power-law three-sublattice order [\textcolor{Red}{a.}] at a nonzero $T$ and [\textcolor{Red}{b.}] in the $T \rightarrow 0$ limit.  The line corresponds to the Markovian random walk value of $\eta_m/2 \equiv 1/8\eta_s$. See Sec.~\ref{Observables} and Sec.~\ref{Results} for a detailed discussion.
	}
\end{figure*}
\begin{figure*}
	\includegraphics[width=\textwidth]{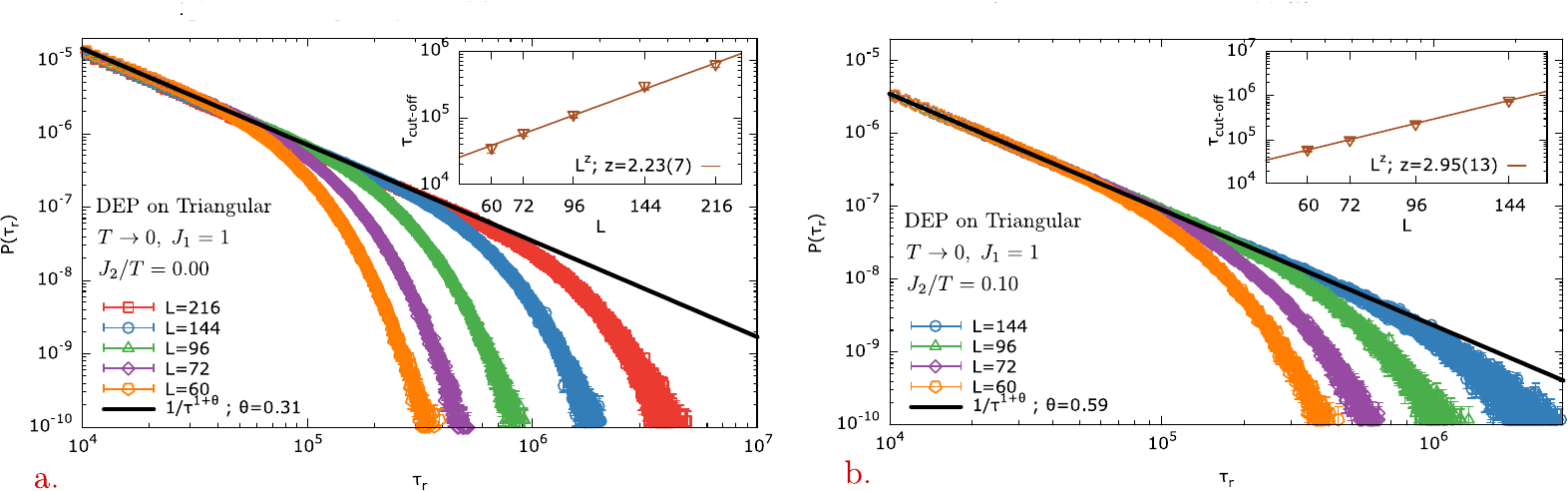}
	\caption{\label{knee} The $L$ dependence of the cutoff scale $\tau_{\rm{cutoff}}$ which cuts off the power-law scaling of $P(\tau_r)$ of the worm length distribution for the DEP algorithm on a periodic $L \times L$ triangular lattice for values of parameters at which the system has power-law three-sublattice order [\textcolor{Red}{a.}] in the $T \rightarrow 0$ limit on a periodic $L \times L$ triangular lattice for [\textcolor{Red}{a.}] $J_2/T=0$ and [\textcolor{Red}{b.}] $J_2/T=0.1$. Fit to a power-law form $\tau_{\rm{cutoff}} \sim L^z$  provides a direct measurement of the dynamical exponent $z$ (inset). See Sec.~\ref{Observables} and Sec.~\ref{Results} for a detailed discussion.
	}
\end{figure*}
\begin{figure}
	\includegraphics[width=\columnwidth]{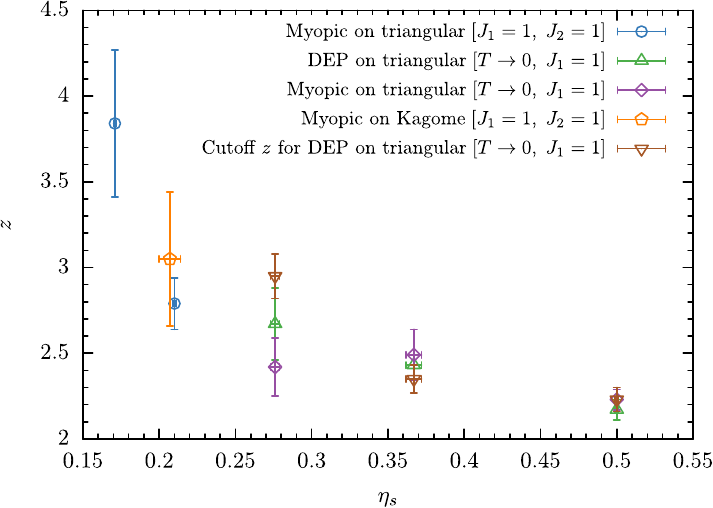}
	\caption{\label{eta_z} Dynamical exponent $z=(2-\eta_m)/(1-\theta)$ extracted from $\eta_m$ (of Fig.~\ref{histdefect}) and $\theta_{\text{measured}}$ (of Fig.~\ref{histvisited}) as a function of $\eta_s$ for all the cases studied here.  Also shown (downward triangles) are the values of $z$ extracted from the finite-size dependence of the cutoff scale $\tau_{\rm{cutoff}}$ (inset of Fig.~\ref{knee}) for the DEP algorithm in the $T \rightarrow 0$ limit. Note that the scaling relation is not used to extract $z$ when $\eta_m > 2$. See Sec.~\ref{Observables} and Sec.~\ref{Results} for a detailed discussion.
	}
\end{figure}

\begin{figure*}
	\includegraphics[width=\textwidth]{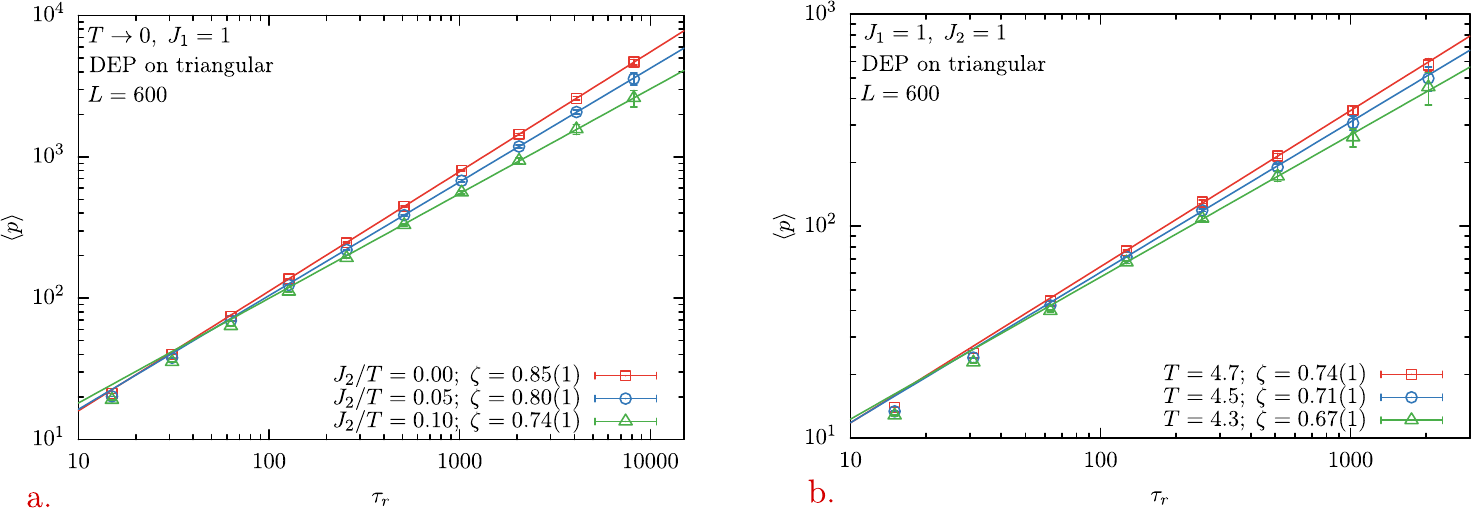}
	\caption{\label{flippedlinks} Average number of dual links $\avg{p}$ flipped by worms of length $\tau_r$ for the DEP worm algorithm on a $L \times L$ triangular lattice for values of parameters at which the system has power-law three-sublattice order [\textcolor{Red}{a.}] in the $T \rightarrow 0$ limit, and [\textcolor{Red}{b.}] at a nonzero $T$. Lines denote fits to a power-law form $\avg{p} \propto {\tau_r}^{\zeta}$. Corresponding plots for the myopic worm algorithm on the trinagular and kagome lattice are shown in Fig.~SM.3 of the Supplemental Material.\cite{suppmat} See Sec.~\ref{Observables} and Sec.~\ref{Results} for a detailed discussion.
	}
\end{figure*}

\begin{figure}
	\includegraphics[width=\columnwidth]{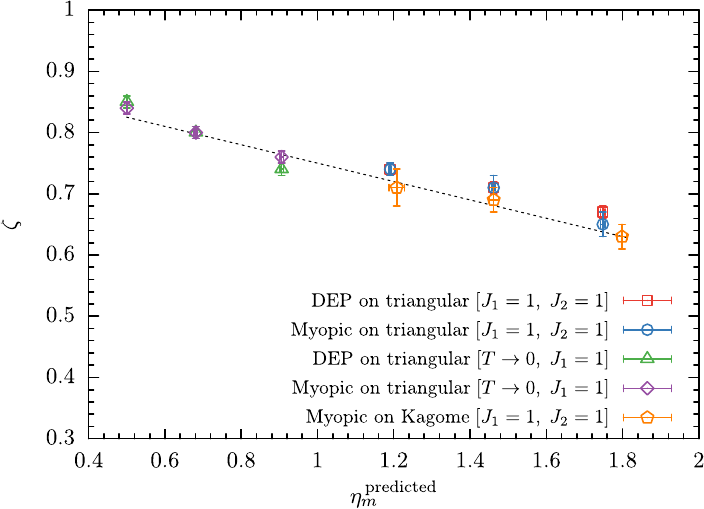}
	\caption{\label{zeta} $\zeta$ extracted from the $\tau_r$ dependence of the average number of flipped dual links $\avg{p}$ (Fig.~\ref{flippedlinks}), plotted  as a function $\eta_m^{\textrm{predicted}}=1/{4 \eta_s}$, where $\eta_s$ is the exponent of the power-law spin-spin correlations at the three-sublattice wavevector. Note that all the points seem to fall on a single trend line, suggesting that $\zeta$ depends in a universal way on $\eta_m$. The dotted line is a linear fit to this dependence, with equation $\zeta = 0.9 - 0.15 \eta_m^{\textrm{predicted}}$. See Sec.~\ref{Observables} and Sec.~\ref{Results} for a detailed discussion.
	}

\end{figure}

\begin{figure*}
	\includegraphics[width=\textwidth]{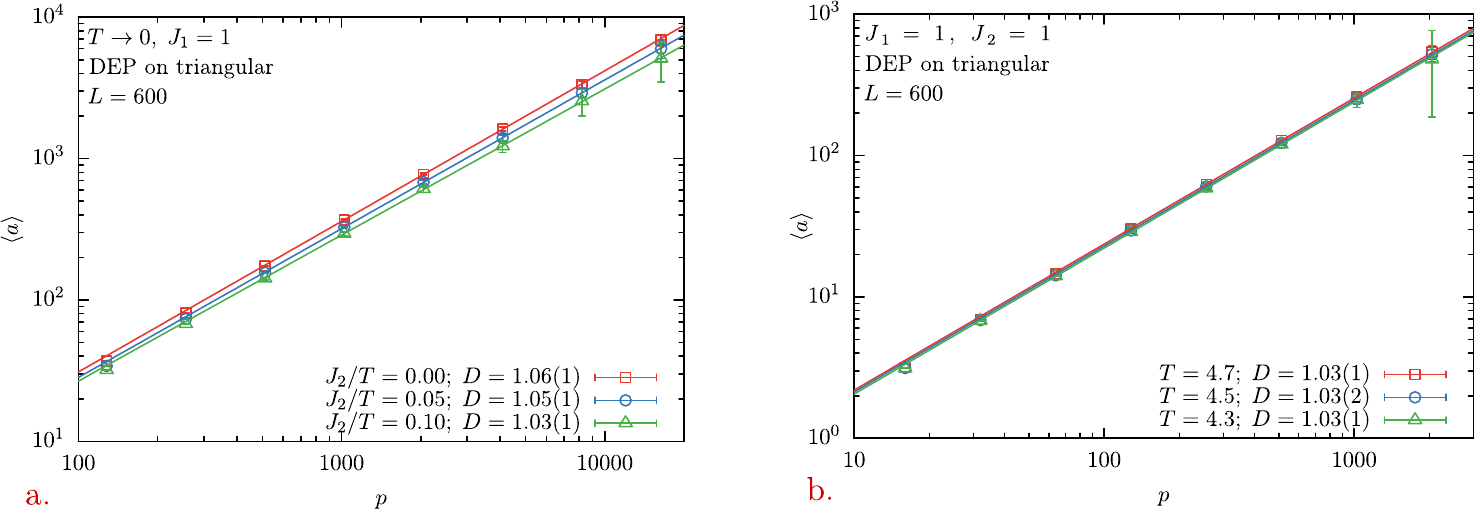}
	\caption{\label{flippedspins} Average number of spins $\avg{a}$ flipped by worms that flip $p$ dual links for the DEP worm algorithm on a $L \times L$ triangular lattice for values of parameters at which the system has power-law three-sublattice order [\textcolor{Red}{a.}] in the $T \rightarrow 0$ limit, and [\textcolor{Red}{b.}] at a nonzero $T$. Lines denote fits to a power-law form $\avg{a}\propto p^{D}$. Corresponding plots for the myopic worm algorithm on the triangular and kagome lattices are shown in Fig.~SM.4 of the Supplemental Material.\cite{suppmat}  See Sec.~\ref{Observables} and Sec.~\ref{Results} for a detailed discussion.
	}
\end{figure*}

\section{Observables}
\label{Observables}
\textbf{Defect-antidefect correlator}: During the worm construction, a defect-antidefect pair is created on the dual lattice, and the antidefect is then moved (keeping the defect fixed) through the dual lattice (in a manner satisfying detailed balance in the enlarged configuration space) until it returns
to the location of the defect and annihilates it, producing a legal dimer configuration that
can be mapped back to a spin configuration. As noted earlier, the defect-antidefect correlator $C_m(\vec{r})$ is proportional to the histogram of the position $\vec{r}$ of the head
relative to the tail of the worm, which can be accumulated during the worm construction.\cite{Alet_Ikhlef_Jacobsen}
We choose a normalization convention where this histogram, when summed over
$\vec{r}$, gives the mean length of worms constructed by the algorithm (in other words,
we measure the number of times the head to tail separation is $\vec{r}$ per worm).
In the power-law three-sublattice ordered phase we expect $C_m(\vec{r}) \sim 1/{r}^{\eta_m}$, with $\eta_m = 1/4\eta_s$. 
During the worm construction, the worm can wind around the torus defined by the periodic boundary conditions used in our study. Even if the worm winds before annihilating, we always record the shortest geometric separation between the head and tail of the worm (modulo the lattice size $L$ in each direction).

\textbf{Worm length or return time distribution}: The number of dual lattice sites (with multiplicities, if a site is visited more than once) visited by the head of the worm during
the worm construction defines the length of the worm, which corresponds in our
random walk analogy to the first-return time of the walk.  A histogram of this gives us $P(\tau_r)$,  the probability distribution of first-return times.

\textbf{Average worm length}: As noted earlier, once our defect-antidefect correlator is normalized to measure the number of times the head to tail separtion is $\vec{r}$ {\em per worm}, then $\langle \tau_r \rangle = \langle v \rangle \equiv \sum_{\vec{r}} C_m(\vec{r})$
In our numerics we measure $\langle v \rangle$, which is expected to scale as $ \sim L^{2-\eta_m}$ in the power-law phase.

\textbf{Average number of flipped links per worm}: When a worm retraces its path, it flips the dimers along the retraced path again, in effect not flipping them in the first place. Thus, counting the number of flipped links is equivalent to measuring the perimeter of the closed path defined by the worm. This closed path is made up of a number of disconnected components in general. As mentioned earlier, this is because every intersection of the worm with its own trace
splits off a closed loop of flipped links. We measure the average number of flipped links per worm $\langle p \rangle$ (summed over all closed components of that path) as a function of the return time $\tau_r$ of the worm.

\textbf{Average number of flipped spins per worm}: After mapping back to the original spin configuration, we can measure the average number of spins on the direct lattice flipped by one worm update. For a simple closed loop, this would be equivalent to measuring the area enclosed by the closed path
defined by the trace of the worm. Since the worm is on a torus, this area can be the either be the inner or outer area with respect to such a simple closed loop. However, since our worm is not a simple closed loop, we do not attempt to define areas in this way. Instead, we start with the final dimer configuration and map it to one of the two spin configurations corresponding to it (we choose one of them randomly, with equal probability). With this in hand, we defne the corresponding 
number of flipped spins to be the smaller of the two numbers for these two choices of final spin configuration. In our measurements, we keep track of the average number of flipped spins $\langle a \rangle$ defined in this way, and study its dependence on the number of flipped links $p$ introduced earlier. 

\section{Results}
\label{Results}

All our measurements are performed on lattice sizes of upto $600 \times 600$ lattice sites for the triangular lattice antiferromagnet and upto $288 \times 288$ unit cells (with three sites per unit cell) for the kagome lattice antiferromagnet. For studying the statistics of worms, we perform one worm update per Monte Carlo step (MCS) and measure all histograms and averages during the worm construction. If the final dimer configuration obtained after the worm construction is not physical from the point of view of the spin model(as explained in Section.~\ref{Algorithms}), we discard the measurements made during the construction of that particular worm. All our data is averaged over $1 \times 10^8$ MCS.

We have performed such measurements in all five cases mentioned in Section.~\ref{Algorithms}: In the $T \rightarrow 0$ limit on the triangular lattice, we study both the DEP and myopic algorithms at three values of $J_2/T$ ($0.00$, $0.05$ and $0.10$), all of which are in the power-law three-sublattice ordered phase. To access the $T>0$ power-law three-sublattice ordered phase, we set $J_1=1$ and $J_2 = +1$. On the triangular lattice, we study both the algorithms in this critical phase at $T=4.3,4.5$ and $4.6$. On the kagome lattice, we study the myopic algorithm in this critical phase at $T=1.24,1.30$ and $1.36$ (all temperatures are measured in units of $J_1=1$).

The defect-antidefect correlator $C_m(\hat{e_x}\frac{L}{s})$ is measured at separation $\vec{r} = \hat{e_x}\frac{L}{s}$ (with $s=2$ for the zero temperature measurements and $s=24$ for the nonzero temperature measurements) on periodic $L\times L$ lattices as a function of lattice size $L$ for $L=288,360,420$ and $600$ on the triangular lattice and $L=96,144,216$ and $288$ on the kagome lattice ($\hat{e}_x$ is one of the Bravais lattice vectors). Fig.~\ref{histdefect}\textcolor{OrangeRed}{(a)} and \ref{histdefect}\textcolor{OrangeRed}{(b)} show this correlator in the $T \rightarrow 0$ limit and at finite $T$ respectively for the DEP worm algorithm on the triangular lattice. The corresponding plots for the myopic worm algorithm on the triangular and kagome lattices are shown in Fig.~SM.1 of the Supplemental Material.\cite{suppmat} In all the above cases we extract $\eta_m$ by fitting a power law to the $L$ dependence of this correlator. We also extract $\eta_m$ from the lattice size $L$ dependence of the average number of dual sites visited per worm (as defined in Section.~\ref{Observables}) using the relation $v /L^2 \sim 1/L^{\eta_m}$. Fig.~\ref{mean_visits}\textcolor{OrangeRed}{(a)} and \ref{mean_visits}\textcolor{OrangeRed}{(b)} show the power-law fits in the $T \rightarrow 0$ limit and finite $T$ respectively for the DEP worm algorithm on the triangular lattice. The corresponding plots for the myopic worm algorithm are shown in Fig.~SM.2 of the Supplemental Material.\cite{suppmat} The value of $\eta_m$ obtained from fits of the mean number of visits matches within error-bars with the value of $\eta_m$ extracted from the defect-antidefect correlator as seen in Fig.~\ref{histdefect}.
In the noninteracting dimer model limit of the dual dimer model ($T \rightarrow 0$ and $J_2/T = 0$), it is well known that $\eta_m= \frac{1}{2}$. Consistent with this, we find $C_m(\hat{e_x}\frac{L}{s}) \sim 1/L^{0.51(1)}$and $\avg{v}/L^2 \sim 1/L^{0.52(2)}$. However, we note that a previous study of a  worm algorithm for the square ice model in the free dimer limit\cite{Barkema_Newman} concluded that $\langle v \rangle \sim L^{1.665(2)}$, which is at odds with what one would expect when $\eta_m =1/2$ (the values of $\eta_m$ and $\eta_d$ are the same for the non-interacting dimer model on the honeycomb and the square lattice). Fig.~\ref{eta_defect} plots the best-fit $\eta_m$ obtained from Fig.~\ref{histdefect} as a function of the spin correlation exponent $\eta_s$ (this exponent is measured by fitting the equilibrium spin correlator at the three sublattice wavevector to a power-law form) for each of these cases. As can be seen, the data agrees very well with the theoretical prediction of $\eta_m^{\textrm{predicted}} = 1/4 \eta_s$ for the $T \rightarrow 0$ case. We note that for $T>0$ cases, the agreement is less impressive but still reasonable.

We measured the probability distribution of worm lengths $\tau_r$ (return times in the random walk language), $P(\tau_r)$, as a function of $\tau_r$ for $L=600$ on the triangular lattice and $L=288$ on the kagome lattice. Fig.~\ref{histvisited}\textcolor{OrangeRed}{(a)} and \ref{histvisited}\textcolor{OrangeRed}{(b)} show the return time distribution in the $T \rightarrow 0$ limit and at finite $T$ respectively for the DEP worm algorithm on the triangular lattice. The corresponding plots for myopic worm algorithm on the triangular and kagome lattices are shown in Fig.~SM.3 of the Supplemental Material.\cite{suppmat} In all these cases, we extract $\theta_\textrm{measured}$ by fitting the probability distribution of return times to a power-law form with exponent $1+\theta$. In these fits we leave out the finite-size effects encountered at large $\tau_r$.

By way of comparison with a more well-known example of worm constructions, we also studied the return time distribution of the worm algorithm for the fully-packed dimer model,\cite{Sandvik_Moessner} on the three-dimensional cubic lattice. In this case, the worm creates a monomer-antimonomer pair, and propagates the antimonomer through the lattice until it recombines with the monomer at the starting site. The monomer-antimonomer correlator on the cubic lattice is controlled by the emergent Coulomb interaction between the monomer and antimonomer. Since this is a power-law potential rather than a logarithmic potential, the effective dimension $d'$ in this case is equal to the spatial dimension: $d'=d=3$. If the dynamical exponent were to take on the usual Markovian random walk value of $z=2$, the return time statistics would be expected to be identical to that of the usual random walk in three dimensions.\cite{Bray} Fig.~\ref{histvisited-dimers} displays a power-law behavior of $P(\tau_r)$  as a function of  $\tau_r$ in this case. The best-fit value $\theta=0.48 \pm0.03$ agrees within errors with the exact value of $1/2$ predicted by Eq.~\ref{Theta} for $d'=3$ and $z=2$. This value of $\theta$ is also consistent with the results for the worm length distributions in Ref.~\onlinecite{Jaubert_Haque_Moessner} for a worm algorithm on the pyrochlore lattice. Thus, in this case, the worm length distributions suggest that correlations between the spatial increments of the random walk renormalize to zero in the long-time limit, yielding a conventional value of $z=2$ for the dynamical exponent.

If the dynamical exponent $z$ were to take on the value $z=2$, then our scaling argument would predict that $\theta=\eta_m/2 \equiv 1/8\eta_s$. We highlight the deviations of the measured value of $\theta$ from this value by plotting $\theta_\textrm{measured}$ as a function of $\eta_m^{\textrm{predicted}}/2 \equiv 1/(8 \eta_s)$ for the finite $T$ and $T \rightarrow 0$ cases in Fig.~\ref{theta}\textcolor{OrangeRed}{(a)} and \ref{theta}\textcolor{OrangeRed}{(b)} respectively. These deviations are evidence that $z \neq 2$. 
Using the scaling relation of Eq.~\ref{scalinglaw}, our results for $\theta$ can be used to obtain the corresponding values of $z$. Independent of this, the value of $z$ can also be determined by a direct measurement of the scale $\tau_{\rm cutoff}(L)$ at which the power-law form of $P(\tau_r)$ is cut off by finite-size effects. This has been illustrated for the DEP algorithm on the triangular lattice in the $T \rightarrow 0$ limit at $J_2/T=0$ and $J_2/T=0.1$ in  Fig.~\ref{knee}\textcolor{OrangeRed}{(a)} and \ref{knee}\textcolor{OrangeRed}{(a)} respectively.

These values of $z$ are seen to match within error-bars with the value of $z$ extracted (using the scaling relation) from $\theta_\textrm{measured}$. This is shown in Fig.~\ref{eta_z} which plots the values of $z$ extracted in various ways for both algorithms in all  the cases studied. From the figure, we see that the value of $z$ appears, within errors, to be determined solely ({\em i.e.} independent of microscopic details like the precise form of the Hamiltonian and the worm construction rules) by the power-law exponent $\eta_s$ that characterises the long-distance behavior of the equilibrium spin correlations. As is clear from this figure, $z$ decreases monotonically with increasing $\eta_s$ and appears to approach the value of $z=2$ in the limit of large $\eta_s$. However, since the largest value of $\eta_s$ accessed in our work is the free-dimer value of $\eta_s=2$, $z > 2$ in the entire regime studied here.

Thus, the worms constructed by these algorithms constitute a particular realization of fractional Brownian motion
with a nontrivial subdiffusive dynamical exponent $z>2$ that is universally determined by the power-law spin correlations of the equilibrium problem. A particular feature of this realization of fractional Brownian motion is the fact that this process has a long-time steady state characterised by the Gibbs distribution for a particle in a logarithmic central potential whose strength is universally determined in terms of the equilibrium defect anti-defect correlation exponent.

We also measured the average number of flipped dual links per worm, $\langle p \rangle$, as a function of the worm length $\tau_r$ for $L=600$ on the triangular lattice and $L=288$ on the kagome lattice. Fig.~\ref{flippedlinks}\textcolor{OrangeRed}{(a)} and \ref{flippedlinks}\textcolor{OrangeRed}{(b)} show this functional dependence in the $T \rightarrow 0$ limit and in the nonzero temperature power-law ordered phase respectively for the DEP worm algorithm on the triangular lattice. The corresponding plots for the myopic worm algorithm on the triangular and kagome lattices is shown in Fig.~SM.4 of the Supplemental Material.\cite{suppmat} In all the cases we find that $\avg{p}$ has a power-law dependence on $\tau_r$: $\langle p \rangle \sim \tau_r^{\zeta}$. The power-law exponent $\zeta$ is shown in Fig.~\ref{zeta} as a function of $\eta_m^{\rm predicted} \equiv 1/4\eta_s$ for each of these five cases. Though we do not have a theoretical prediction for this dependence, we note that all the measured data points seem to fall on a single curve, as would be expected if the geometric properties of the worms were universally determined by the long-distance behaviour of equilibrium correlations. Since this universal dependence appears approximately linear in the range studied, we fit it to a straight line, obtaining: $\zeta \approx 0.9-0.15\eta_m$

In addition, we measured the average number of flipped spins per worm on the direct lattice $\langle a \rangle$ as a function of $p$, the number of flipped dual links, for $L=600$ on the triangular lattice and $L=288$ on the kagome lattice. In measuring this quantity, we exploit the fact that the final dimer configuration obtained after the construction of one worm corresponds to two spin configurations related by a global spin flip. Keeping this in mind, we compare the number of flipped spins corresponding to both these final spin configurations, and record the smaller of these two numbers. Fig.~\ref{flippedspins}\textcolor{OrangeRed}{(a)} and \ref{flippedspins}\textcolor{OrangeRed}{(b)} show the distribution in the $T \rightarrow 0$ limit and at finite $T$ for the DEP worm algorithm on the triangular lattice. The corresponding plots for myopic worm algorithm on the triangular and kagome lattice are shown in Fig.~SM.5 of the Supplemental Material.\cite{suppmat} In all the above cases we extract the exponent $D$ by fitting this functional dependence to a power law form. For worms that do not intersect themselves before closing, this would amount to plotting the enclosed area as a function of perimeter of the worm. However, we caution that the exponent $D$ is {\em not} the fractal dimension of the cluster constructed by the worm, since this would involve the radius of gyration, rather than the perimeter. When we perform the fits, we find that the measured exponent $D \approx 1$ in all five cases studied. To understand this better, we have looked at the actual traces of the worms in all cases, and found that the worms defined by these algorithms intersect themselves very often. The spin cluster being flipped thus consists of many small components, which correspond roughly to the ``interiors'' of each of the daughter loops discussed in Sec.~\ref{Random_walker}, and the area of these individual components does not scale with the measured total perimeter. For such worms, it is quite natural that the total perimeter and the total number of flipped spins scale in the same way, {\em i.e.} $D \approx 1$. Additionally, although we have not tried to quantify this aspect of the random geometry of our worms, we note that ramified fractal clusters can also quite generically have a perimeter that scales as the area.\cite{Leath}

\section{Outlook}
\label{Outlook}
Our results imply that the worms studied here define a discrete-time realization of a fractional Brownian motion
which has a conventional steady-state given by the equilibrium Gibbs distribution of
a particle in a logarithmic central attractive potential. The dynamics of these worms is non-Markovian because time steps are correlated with each other via their dependence on the power-law correlated background dimer liquid.
Stochastic
equations related to such non-Markovian processes with correlated steps have been studied for some time now. Ageing and steady-state behaviour of solutions to
such equations, particularly in the presence of a confining potential, have also
been of interest.\cite{Metzler,Stationary_fBM} It would
therefore be interesting to ask if a continuous-time stochastic equation of this type emerges
as the correct description of some scaling limit of the worm construction process studied here. In this regard, a promising line of investigation would be to use the
well-understood continuum effective field theory formulation of the non-interacting
honeycomb lattice dimer model to formulate an appropriate stochastic differential
equation that captures the large-distance long-time properties of the worm
construction.

In our work, we focused on the properties of the worms. As already emphasized in Sec.~\ref{Random_walker}, one can also study the
properties of the overlap loops formed by superposing the updated dimer configuration
on the original dimer configuration. These are closely related to the individual components
of the trace of the worm (since each self-intersection splits off a component). In future
work, it would be interesting to study the statistics of these daughter loops
obtained from a single worm, and compare these statistical properties with the known properties of the ensemble of overlap loops obtained by superposing two dimer configurations drawn independently of each other from the equilibrium dimer ensemble.
Finally, a similar picture for the worm-length distribution is possible in other applications of worm algorithms to two-dimensional critical points/phases, and it would be interesting to study
the values of dynamical exponent $z$ and persistence exponent $\theta$ associated with these worm constructions.

\section{Acknowledgments} We acknowledge useful discussions with Fabien Alet and
Pranay Patil on their own unpublished observations about worm statistics of similar
worm algorithms. Our computational
work relied on the computational facilities of the Dept. of Theoretical Physics
of the Tata Institute of Fundamental Research (TIFR). GR was supported
by a graduate fellowship at the Tata Institute of Fundamental Research during the first part of this work, which formed
a part of his Ph.D thesis at the TIFR, and by a postdoctoral fellowship at the Okinawa Institute of Science and Technology during the completion of this work. The work
of DD was supported in part by a J. C. Bose Fellowship of SERB, DST India (DST-SR-S2/JCB-24/2005). The work of KD at TIFR was supported by DAE India, and in part by a J. C. Bose Fellowship of SERB, DST India (JCB/2020/000047) and the Infosys Foundation under the aegis of the Infosys-Chandrasekharan Random Geometry Center.

\appendix*

\begin{widetext}

\newpage

  \section{Suplemental Material for ``Fractional Brownian motion of worms in worm algorithms for frustrated Ising magnets''}

  \setcounter{figure}{0} \renewcommand{\thefigure}{SM.\arabic{figure}}

    This Supplemental Material provides additional supporting evidence for the numerical results presented in the main text. Here we focus on displaying additional data and analysis regarding  the geometry of worms made by the myopic worm algorithm in the nonzero temperature phase of the Ising antiferromagnet with power-law three-sublattice order on the kagome lattice, as well as corresponding details of the behaviour of worms made by the myopic worm algorithm on the triangular lattice, both in the $T \to 0$ limit, and in the nonzero temperature power-law three-sublattice ordered phase. Since this data and the associated analysis is entirely analogous to that presented in the main text for the DEP algorithm on the triangular lattice, we confine ourselves to displaying here the relevant data and its analysis in five figures with captions that should be read in conjunction with the captions of the corresponding figures in the main text.

\begin{figure}[h]
	\includegraphics[width=\textwidth]{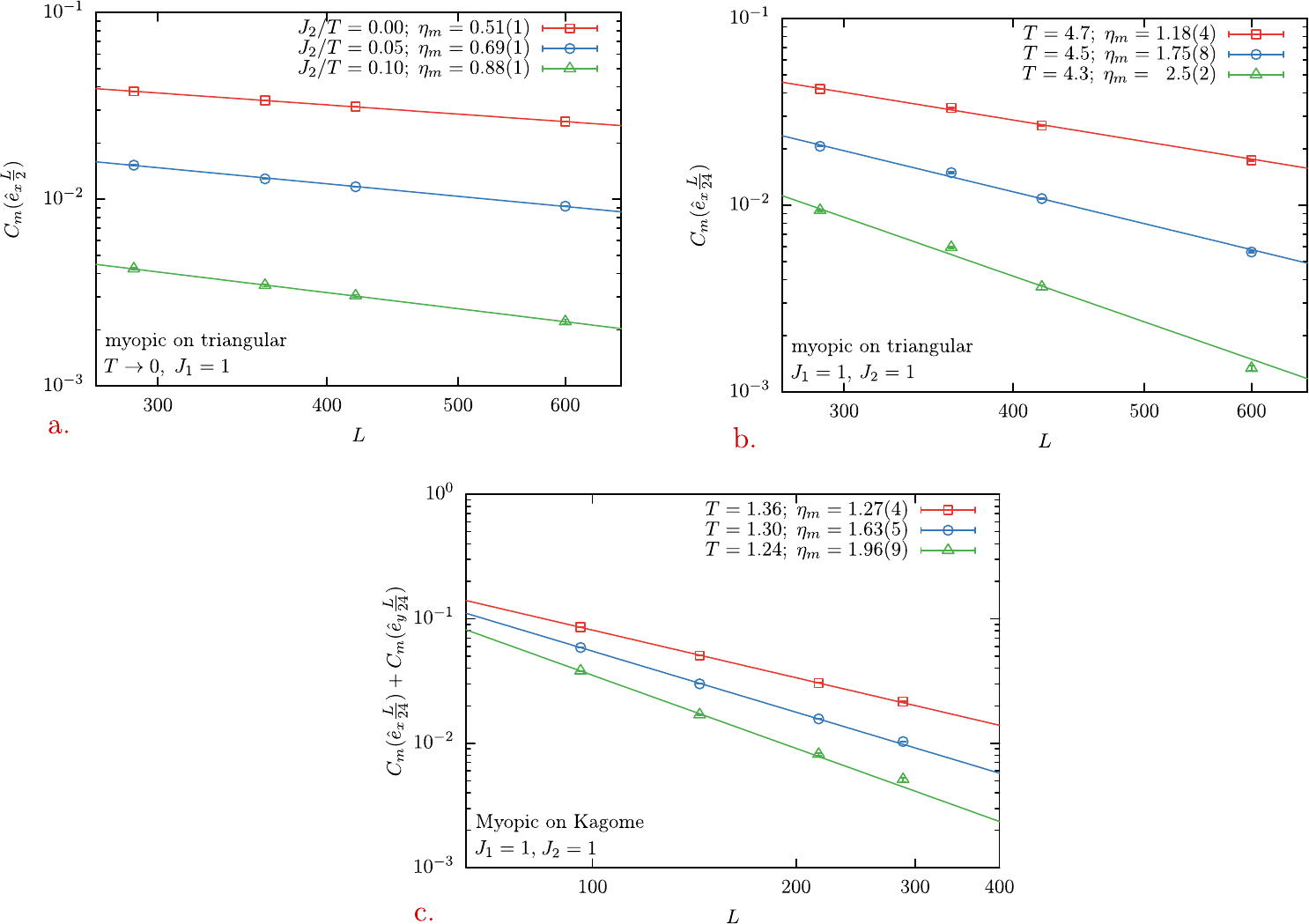}
	\caption{\label{supp-histdefect}The lattice size $L$ dependence of the defect-antidefect correlator $C_{m} \left( \hat{e_x} \frac{L}{a}  \right)$ at separation $\hat{e_x} \frac{L}{a}$ for the myopic worm algorithm on a periodic $L \times L$ [\textcolor{Red}{a.}] triangular lattice in the $T \rightarrow 0$ limit, [\textcolor{Red}{b.}] triangular lattice at a nonzero $T$ and [\textcolor{Red}{c.}] kagome lattice at a nonzero $T$ for values of parameters at which the system has power-law three-sublattice order. Lines denote fits to a power-law form $C_{m} \left( \hat{e_x} \frac{L}{a}  \right) \propto 1/L^{\eta_m}$.  
	}
\end{figure}
\begin{figure}[t]
	\includegraphics[width=\columnwidth]{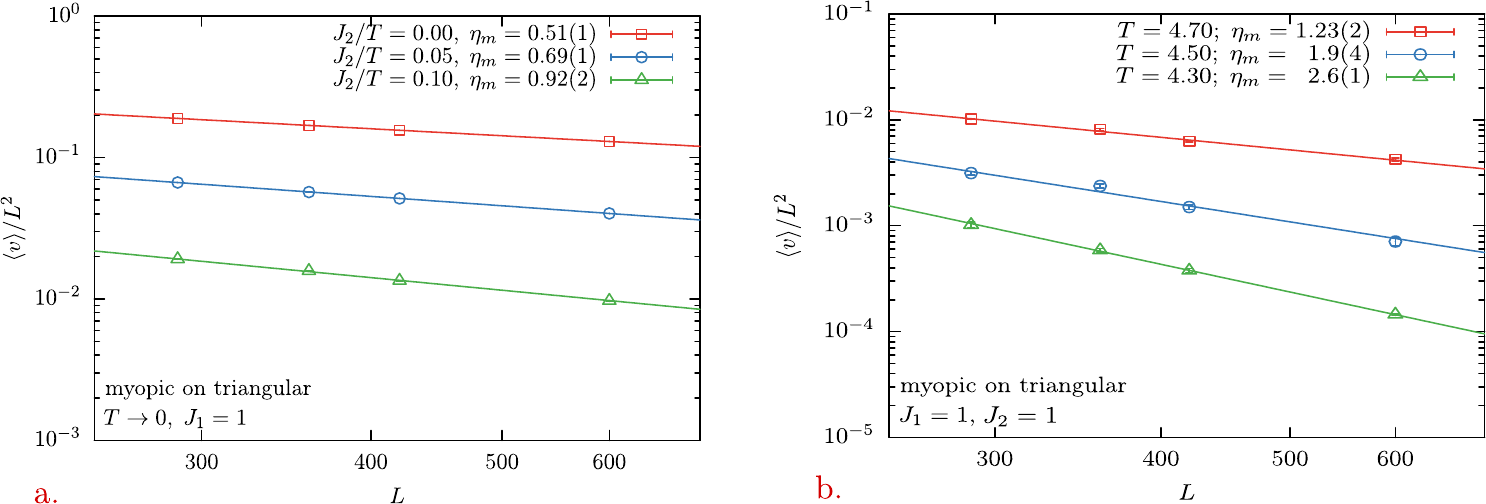}
	\caption{\label{supp-mean_visits}The lattice size $L$ dependence of the average number of dual lattice sites visited per worm, $\avg{v}$, for the myopic worm algorithm on a periodic $L \times L$ [\textcolor{Red}{a.}] triangular lattice in the $T \rightarrow 0$ limit and  [\textcolor{Red}{b.}] triangular lattice at a nonzero $T$ for values of parameters at which the system has power-law three-sublattice order. Since $\avg{v}/L^2 \sim 1/L^{\eta_m}$, the power-law fits give us an alternate measurement of $\eta_m$.
	}
\end{figure}
\begin{figure}[t]
	\includegraphics[width=\textwidth]{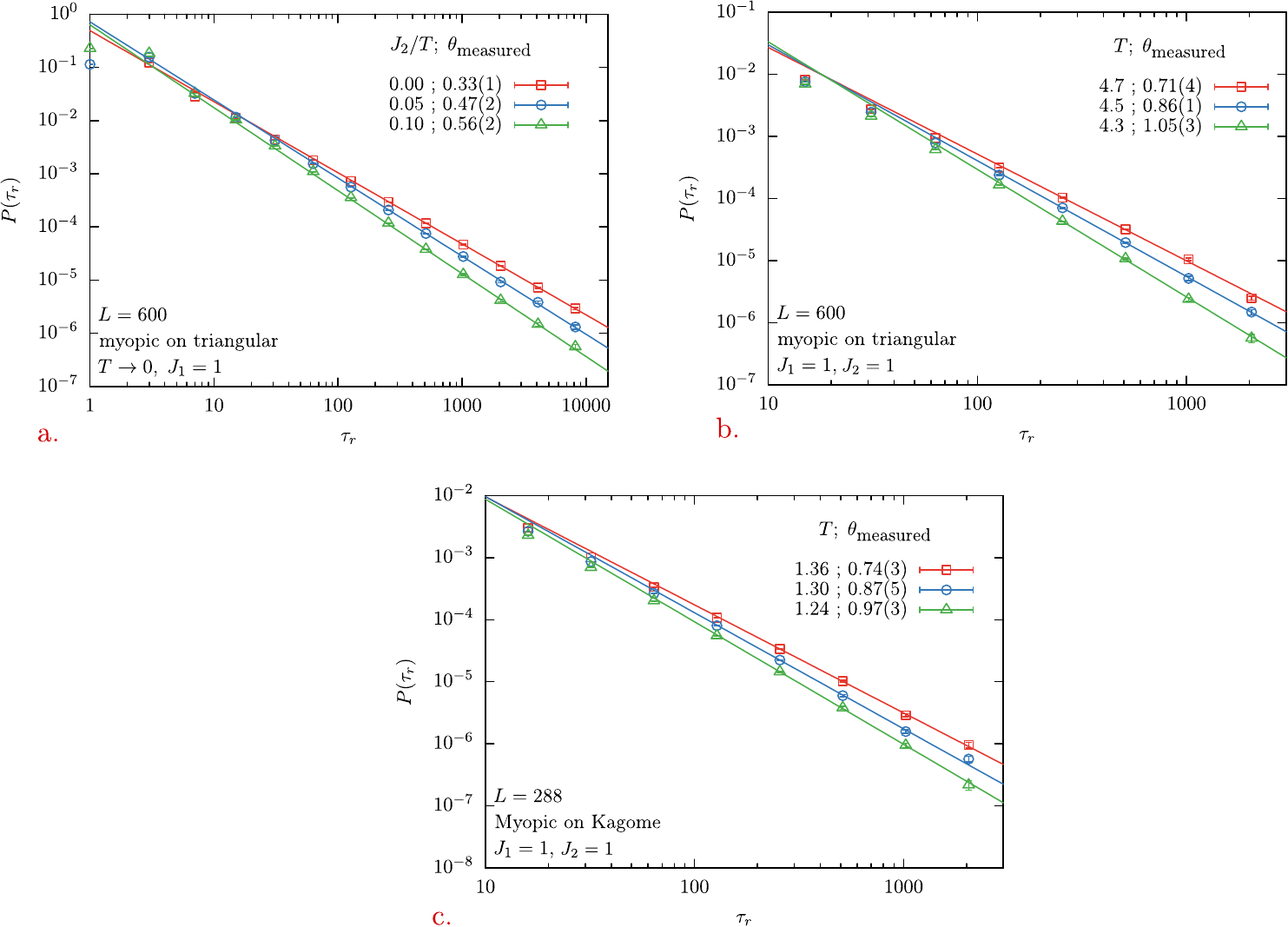}
	\caption{\label{supp-histvisited} Probability distribution $P(\tau_r)$ of the worm length $\tau_r$, {\it i.e.} the number of sites $\tau_r$ of the dual lattice visited by a worm of the myopic worm algorithm on a periodic $L \times L$ [\textcolor{Red}{a.}] triangular lattice in the $T \rightarrow 0$ limit, [\textcolor{Red}{b.}] triangular lattice at a nonzero $T$ and [\textcolor{Red}{c.}] kagome lattice at a nonzero $T$ for values of parameters at which the system has power-law three-sublattice order. Lines denote fit to a power-law form $P(\tau_r) \propto 1/{\tau_r}^{1+\theta_{\text{measured}}}$.
	}
\end{figure}
\begin{figure}[t]
	\includegraphics[width=\textwidth]{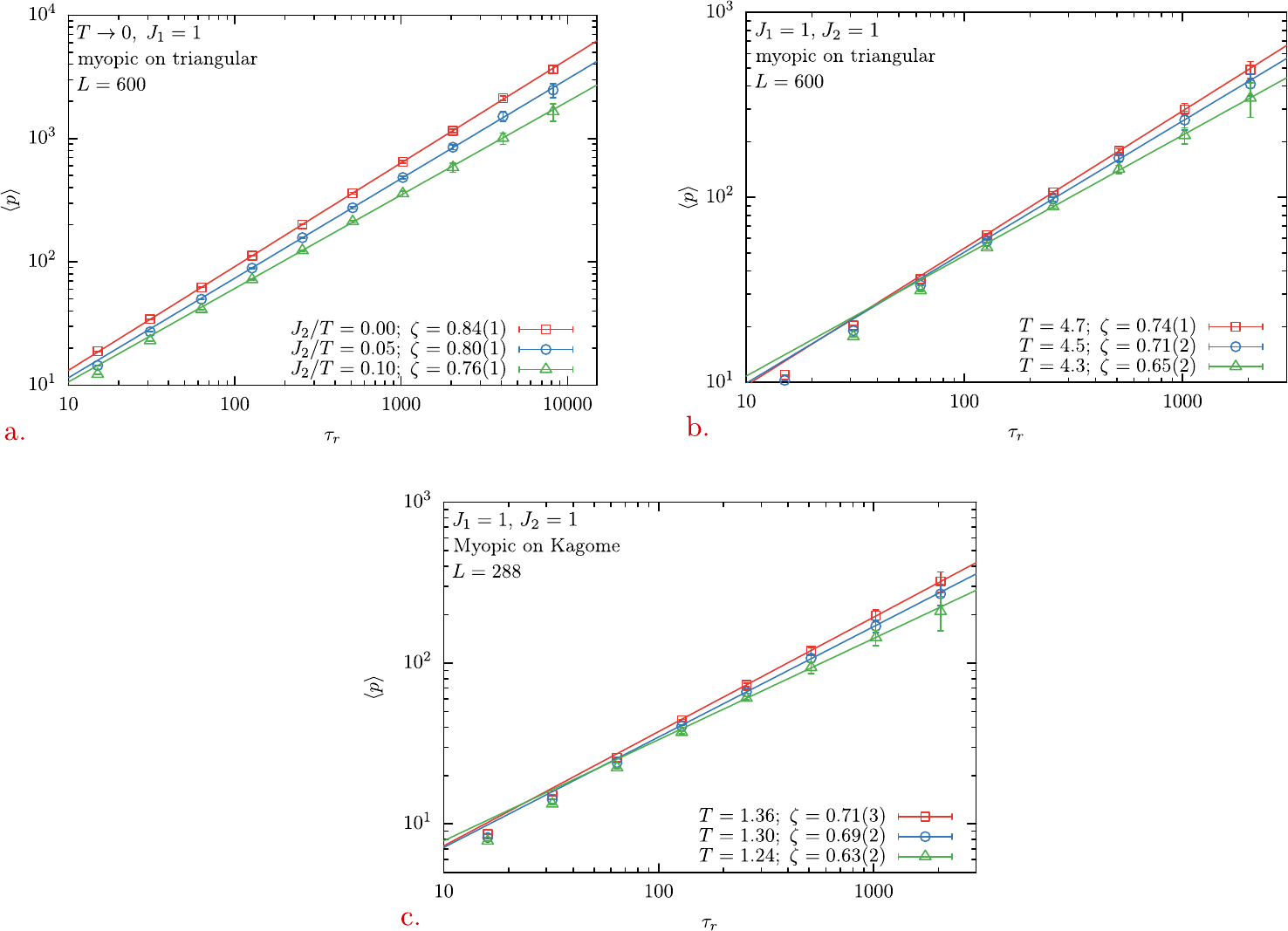}
	\caption{\label{supp-flippedlinks} Average number of dual links $\avg{p}$ flipped by worms of length $\tau_r$ for the myopic worm algorithm on a periodic $L \times L$ [\textcolor{Red}{a.}] triangular lattice in the $T \rightarrow 0$ limit, [\textcolor{Red}{b.}] triangular lattice at a nonzero $T$ and [\textcolor{Red}{c.}] kagome lattice at a nonzero $T$ for values of parameters at which the system has power-law three-sublattice order. Lines denote fits to a power-law form $\avg{p} \propto {\tau_r}^{\zeta}$. 
	}
\end{figure}
\begin{figure}[h]
	\includegraphics[width=\columnwidth]{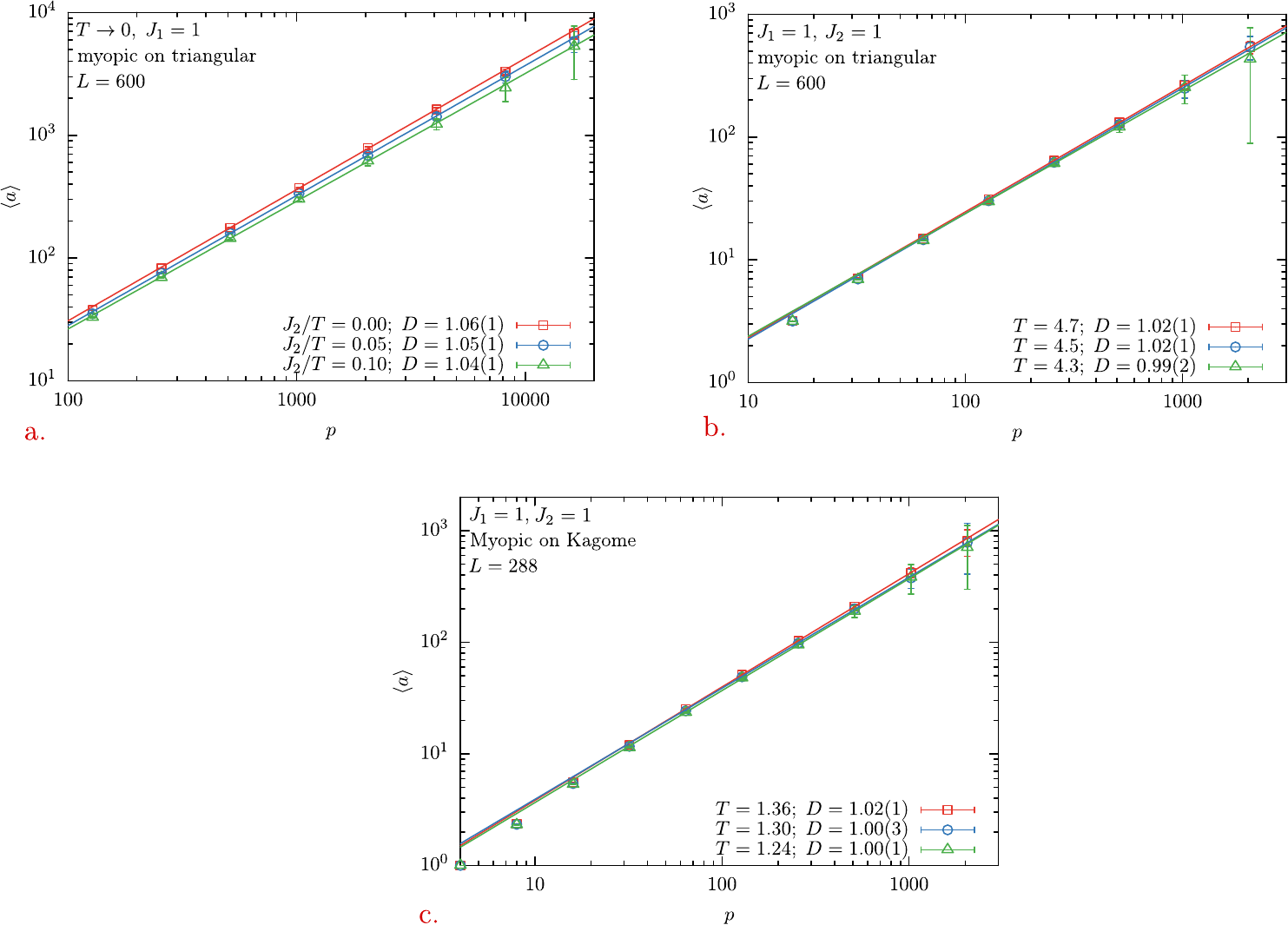}
	\caption{\label{supp-flippedspins}Average number of spins $\avg{a}$ flipped by worms that flip $p$ dual links for the myopic worm algorithm on a periodic $L \times L$ [\textcolor{Red}{a.}] triangular lattice in the $T \rightarrow 0$ limit, [\textcolor{Red}{b.}] triangular lattice at a non-zero $T$ and [\textcolor{Red}{c.}] kagome lattice at a nonzero $T$ for values of parameters at which the system has power-law three-sublattice order. Lines denote fits to a power-law form $\avg{a} \propto p^{D}$.
	}
\end{figure}

\end{widetext}

\end{document}